\documentclass[aps,pra,twocolumn,showpacs,amsmath,amssymb,amsfonts,superscriptaddress]{revtex4-1}
\usepackage{graphicx}
\usepackage{dcolumn}
\usepackage{bm}
\usepackage{color}
\usepackage{multirow}
\usepackage{comment}
\usepackage{wasysym}
\usepackage{marvosym}

\definecolor{purple}{rgb}{0.57, 0.36, 0.51}
\definecolor{dgreen}{rgb}{0.42, 0.56, 0.14}
\definecolor{hotmagenta}{rgb}{1.0, 0.11, 0.81}

\renewcommand{\P}{{\mathcal P}}
\renewcommand{\L}{{\mathcal L}}
\renewcommand{\S}{{\mathcal S}}

\begin{document}
\author{Katarzyna Macieszczak}
\affiliation{School of Physics and Astronomy, The University of Nottingham, Nottingham, NG7 2RD, United Kingdom}
\affiliation{Centre for the Mathematics and Theoretical Physics of Quantum Non-equilibrium Systems, University of Nottingham, Nottingham NG7 2RD, UK}
\author{YanLi Zhou}
\affiliation{College of Science, National University of Defense Technology, Changsha, 410073, China}
\author{Sebastian Hofferberth}
\affiliation{Department of Physics, Chemistry and Pharmacy,
University of Southern Denmark, Odense, Denmark}
\author{Juan P. Garrahan}
\author{Weibin Li}
\author{Igor Lesanovsky}
\affiliation{School of Physics and Astronomy, The University of Nottingham, Nottingham, NG7 2RD, United Kingdom}
\affiliation{Centre for the Mathematics and Theoretical Physics of Quantum Non-equilibrium Systems, University of Nottingham, Nottingham NG7 2RD, UK}

\title{Metastable decoherence-free subspaces and electromagnetically induced transparency in interacting many-body systems}

\begin{abstract}
We investigate the dynamics of a generic interacting many-body system under conditions of electromagnetically induced transparency (EIT). This problem is of current relevance due to its connection to non-linear optical media realized by Rydberg atoms. In an interacting system the structure of the dynamics and the approach to the stationary state becomes far more complex than in the case of conventional EIT. In particular, we discuss the emergence of a metastable decoherence free subspace, whose dimension for a single Rydberg excitation grows linearly in the number of atoms. On approach to stationarity this leads to a slow dynamics which renders the typical assumption of fast relaxation invalid. We derive analytically the effective non-equilibrium dynamics in the decoherence free subspace which features coherent and dissipative two-body interactions. We discuss the use of this scenario for the preparation of collective entangled dark states and the realization of general unitary dynamics within the spin-wave subspace.
\end{abstract}


\maketitle


\section{Introduction} The phenomenon of electromagnetically induced transparency (EIT) is currently extensively studied both theoretically and experimentally~\cite{fleischhauer_electromagnetically_2005,Hofferberth2016d,murray_review_2016}. It finds applications in the context of quantum memories~\cite{li_quantum_2016,distante_storing_2017} and slow light~\cite{boller_observation_1991} as well as in the mediation of effective photon-photon interactions~\cite{Adams2010,Adams2013,Grangier2012,Grangier2012,Kuzmich2012b,Kuzmich2013,Vuletic2012,Vuletic2013b}. These are important ingredients for optical quantum computing and permit the creation of non-linear optical elements such as single-photon switches and transistors ~\cite{Baur_photon_switch,gorniaczyk_single-photon_2014,tiarks_photon_transistor_2014,gorniaczyk_enhancement_2016,li_coherence_2015,murray_many-body_2016}, single-photon absorbers~\cite{tresp_single-photon_2016},
as well as photon gates~\cite{gorshkov_photon-photon_2011}.

In the context of EIT a common assumption is made concerning a separation of timescales between the slow propagation of the optical fields and the fast dynamics of the atomic medium. The latter is therefore assumed to be always in its stationary state and the transmission properties of the atomic medium are determined by the corresponding stationary state density matrix.
\begin{figure*}[ht!]
\begin{center}
\includegraphics[width=\textwidth]{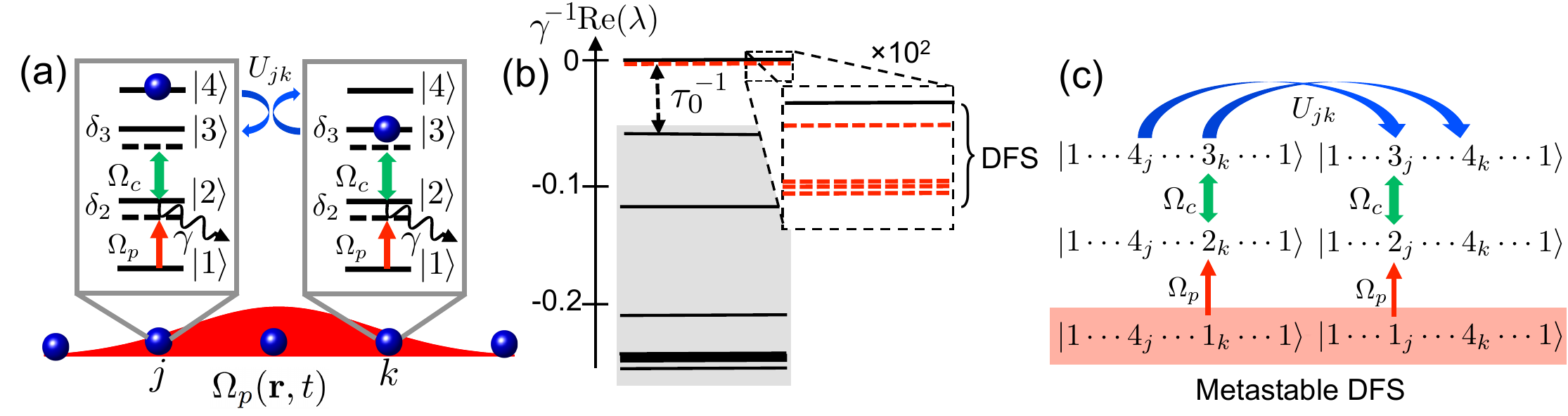}
\vspace*{-7mm}
\caption{
{\bf Metastable DFS of an $N$-atom system:}
(a) Level scheme and transitions. (b) Spectrum of the master operator $\mathcal{L}$ displaying a separation of eigenvalues between low-lying modes $(\lambda_{1}=0,\lambda_2,...,\lambda_{N^2})$ (full and dashed) corresponding to the long-time dynamics in Eq.~\eqref{eq:Leff}, and fast modes $\lambda_{k>N^2}$ (shaded). Data for $N=3$ atoms with van der Waals interactions on a one-dimensional lattice (lattice spacing $a$),  dispersion coefficients $C_{34} = 1.3\times \gamma a^6$ and  $C_{\text{ex}} = \gamma a^6$, in the presence of uniform fields $\Omega_{p}=\Omega_c/50=\gamma/50$ and the detuning $\delta_2=\delta_3=0$.  (c) Exchange interactions $U_{jk}$ together with the probe-field coupling lead to a slow non-local dynamics within the $N$-dimensional DFS of spin-waves, and render it metastable. 
}\vspace*{-5mm}
\label{fig:1}
\end{center}
\end{figure*}

In this work we show, however, that the dynamics of an interacting atomic medium is not necessarily fast due to the emergence of quantum metastability~\cite{macieszczak_metastability_2016}. To illustrate this we explore analytically the long-time evolution of a generic interacting many-body ensemble under EIT conditions. We demonstrate that the effective dynamics in fact takes place within a metastable decoherence free subspace (DFS) of spin-waves (SWs), and features both dissipative and coherent two-body interactions~\cite{coherent_dissipative_interaction}. The emerging slow timescales lead to a violation of the typical assumption of fast system relaxation~\cite{fleischhauer_electromagnetically_2005} which drastically affects the system's (non-local) optical response. We show analytically that the effective long-time dynamics can be employed for preparing stationary pure and entangled SW dark states~\cite{Arimondo1996,Harris1997,diehl_quantum_2008,kraus_preparation_2008}, and, in the limit of weak interactions, allows to implement arbitrary unitary evolution within the metastable DFS~\cite{zanardi_coherent_2014,zanardi_geometry_2015}. Our study is relevant for recent investigations in the context of Rydberg quantum optics, but more generally sheds light on non-trivial effects due to quantum metastability in interacting many-body systems.

The structure of the paper is as follows. In Sec.~\ref{sec:system} we introduce a generic interacting many-body system in EIT configuration. In Sec.~\ref{sec:MM} we discuss its metastable states and in Sec.~\ref{sec:Leff} derive the effective long-time dynamics. In Sec.~\ref{sec:pol} we study the optical response of the system. Finally, in Sec.~\ref{sec:rhoss} we discuss stationary states of the effective dynamics and show how the dynamics can be used for pure entangled state preparation, as well as realisation of universal unitary gates (Sec.~\ref{sec:unitary}). The results of the paper are summarized in Sec.~\ref{sec:summary}.

\section{The system~\label{sec:system}} We consider the dynamics of a system of $N$ interacting atoms with four relevant electronic levels, as depicted in Fig.~\ref{fig:1}a: the ground state $|1\rangle$, a low-lying short-lived excited state $|2\rangle$ and two long-lived states $|3\rangle$ and $|4\rangle$. The $|1\rangle\leftrightarrow|2\rangle$-transition is driven by a (weak) probe field with Rabi frequency $\Omega_{p}(\mathbf{r},t)$, while the $|2\rangle\leftrightarrow |3\rangle$-transition is coupled by a (strong) control laser field with Rabi frequency $\Omega_{c}(\mathbf{r})$, giving rise to a typical EIT configuration. Within the dipole and rotating wave approximations, the laser-atom coupling Hamiltonian is
\begin{align}
\!H_j=\!\!\sum_{n=2,3}\!\!\delta_n\sigma_{nn}^{j}\!
+\left(\Omega_p(\mathbf{r}_j,t)\sigma_{21}^{j}+\Omega_c(\mathbf{r}_j)\sigma_{32}^{j}
\!+\text{h.c.}\right)\!,\!\label{eq:Ham}
\end{align}
where $\sigma_{ab}^{j}=|a_j\rangle\!\langle b_j|$ and $\delta_n (n=2,\,3)$ are the detunings of the respective lasers.

Atoms interact via the density-density interaction 
\begin{equation}
V_{jk}= \sum_{n,m=3,4}V_{jk}^{nm}\, \sigma_{nn}^{j}\sigma_{mm}^{k}
\end{equation}
and the exchange interaction 
\begin{equation}
U_{jk}= U^{34}_{jk}\,\sigma_{43}^{j}\sigma_{34}^{k}+\mathrm{h.c.}.
\end{equation}
Interactions among the low-lying states, $|1\rangle$ and $|2\rangle$, are neglected. This choice is rather generic, but also motivated by recent investigations of EIT within Rydberg gases~\cite{tiarks_photon_transistor_2014,gorniaczyk_single-photon_2014,gorniaczyk_enhancement_2016,tresp_single-photon_2016,tiarks_optical_2016,thomson_symmetry_2017}, where the upper states $|3\rangle$ and $|4\rangle$ correspond to two different Rydberg levels: two Rydberg atoms, located at positions $\mathbf{r}_j$ and $\mathbf{r}_k$, interact via van der Waals interaction $V_{jk}^{mn}=C_{mn}/|\mathbf{r}_{j}-\mathbf{r}_{k}|^6$, and exchange interaction $U_{jk}^{34}=C_{\mathrm{ex}}/|\mathbf{r}_{j}-\mathbf{r}_{k}|^6$, with dispersion coefficients $C_{\textrm{ex}}$, $C_{33}$, $C_{44}$, and $C_{34}=C_{43}$~\cite{li_electromagnetically_2014,coherence_Li_2015,thomson_symmetry_2017}.

Coherent dynamics is supplemented by dissipative decay of the short-lived state $|2\rangle$ into the ground state $|1\rangle$ at rate $\gamma$. The evolution of the density matrix $\rho$ is governed by a quantum master equation, with master operator $\mathcal{L}$, given by~\cite{Lindblad1976,Gorini1976}
\begin{equation}
\frac{\mathrm{d}}{\mathrm{d}t}\rho=\mathcal{L}\,\rho=-i\left[H,\rho\right]+\gamma\sum_{j=1}^N\mathcal{D}(\sigma_{12}^{j})\,\rho, \label{eq:master}
\end{equation}
with  the dissipator $\mathcal{D}(L)\rho:=L\rho L^\dagger-\frac{1}{2}\{L^\dagger L,\rho\}$ and the Hamiltonian $
H=\sum_{j=1}^N H_j + \sum_{j>k}^N[U_{jk}+V_{jk}]$.

\section{Metastable manifolds~\label{sec:MM}}
\indent{\it Metastable manifolds in non-interacting EIT}. 
It is instructive to first consider the non-interacting case in order to get an idea of the resulting decoherence free subspace (DFS), the emergence of metastability and the corresponding timescales. In the absence of interactions state $|4\rangle$ is dynamically disconnected from the remaining levels, cf. Fig.~\ref{fig:1}a, and each atom possesses two stationary states: a mixed state $\rho_{\mathrm{ss},\mathbf{r}}$ supported on the lower three levels, and the pure non-decaying excited state $|4\rangle\!\langle 4|$.

An interesting situation occurs when both stationary states are pure, i.e. $\rho_{\mathrm{ss},\mathbf{r}}\mapsto|\psi_\mathbf{r}\rangle\!\langle\psi_\mathbf{r}|$. It follows that $|\psi_\mathbf{r}\rangle$ is a so-called dark state, i.e. $\sigma_{12} |\psi_\mathbf{r}\rangle=0$, and also an eigenstate of the local Hamiltonian, $H_{\mathbf{r}} |\psi_\mathbf{r}\rangle=E_\mathbf{r}|\psi_\mathbf{r}\rangle$~\cite{diehl_quantum_2008,kraus_preparation_2008}. Therefore, also coherences between $|\psi_\mathbf{r}\rangle$ and $|4\rangle$ become stationary, so that $|\psi_\mathbf{r}\rangle$ and $|4\rangle$ span a DFS. On resonance, i.e. $\delta_3=0$, we have
\begin{equation}
|\psi_\mathbf{r}\rangle= \frac{\Omega_c^*(\mathbf{r})|1\rangle-\Omega_p(\mathbf{r})|3\rangle}{\sqrt{\lvert\Omega_c(\mathbf{r})\rvert^2+\lvert\Omega_p(\mathbf{r})\rvert^2}},\label{eq:dark}
\end{equation} 
and the dark stationary state is reached on a timescale $\tau_0$, determined by the spectral gap of the master operator $\mathcal{L}$, cf. Fig.~\ref{fig:1}b. For small non-zero detuning $\delta_3$, the single-atom DFS, spanned by $|\psi_\mathbf{r}\rangle$ and $|4\rangle$, is no longer truly stationary, but becomes metastable~\cite{macieszczak_metastability_2016}, as the stationary state degeneracy is partially lifted and low-lying slow modes appear in the spectrum of $\mathcal{L}$, see Fig.~\ref{fig:1}b. These modes govern the long-time dynamics within the DFS at $t\gtrsim \tau =\mathcal{O}[1/(\delta_3^{2}\tau_0)]$, which relaxes the system to the actual stationary state, i.e. a mixture of $\rho_{\mathrm{ss},\mathbf{r}}$ and $|4\rangle\!\langle 4|$~\footnote{As $\delta_3\neq0$ perturbs a DFS, the long-time dynamics timescale $\tau$ is actually determined with non-dissipative relaxation time $\tau_0'\leq\tau_0$ given by the inverse of the imaginary gap of the effective Hamiltonian of the single-atom dynamics $\mathcal{L}$ at $\delta_3=0$, i.e. $H_j-\frac{i}{2}\sigma_{22}^{j}$ instead of $\tau_0$~\cite{albert_adiabatic_2016}, see Appendix~\ref{app:NI}.}.

When using EIT to control light fields, e.g. for quantum memories or light storage, the response of the atomic ensemble to the incident field $\Omega_p(\mathbf{r}_j,t)$ is determined by the coherence between the low-lying states $|1\rangle$ and $|2\rangle$, $\langle  \sigma_{12}\rangle_\mathrm{ss}$, calculated in the stationary state $\rho_{\mathrm{ss},\mathbf{r}}$ \cite{fleischhauer_electromagnetically_2005}. This implicitly assumes that the timescale $\tau_p$ connected to the probe field dynamics, is significantly longer than the relaxation time $\tau_0$ of the atomic ensemble, defining an adiabaticity condition. At resonance, $\delta_3=0$, the ensemble then remains in the dark state $|\psi_\mathbf{r}\rangle$ by adiabatically following $\Omega_p(\mathbf{r},t)$, therefore $\langle  \sigma_{12}\rangle_\mathrm{ss}=0$, leading to transparency of the ensemble~\cite{fleischhauer_electromagnetically_2005}. 
However, one might wonder why the adiabaticity condition can be met also in the non-resonant case, $\delta_3\neq 0$, where the relaxation time $\tau$ can in principle become arbitrarily long. The answer is that the coherence relaxes to its stationary value, $\langle  \sigma_{12}\rangle_\mathrm{ss}$,  at the fast timescale $\tau_0\ll\tau_p$, as the slow dynamics related to $\tau$ corresponds to dephasing of coherences between $|\psi_\mathbf{r}\rangle$ and $|4\rangle$, which are irrelevant for the optical response. Thus, although the non-interacting system for $\delta_3\neq0$ becomes in fact metastable,  this does not invalidate the adiabaticity condition.\\

\indent{\it Metastable manifolds of the interacting system}. In the presence of interactions, in particular the long-range exchange interaction $U_{jk}$, this no longer holds true. Here a slow collective dynamics within a metastable DFS emerges which affects the optical response by introducing a new timescale that becomes relevant for EIT.

To study this in detail, we note that in the absence of the probe field, the ground level $|1\rangle$ is disconnected from the dynamics, as shown in Fig.~\ref{fig:1}a. For a single atom also the level $|4\rangle$ is disconnected.  For many atoms this is no longer the case, but the exchange and density-density interactions connect $|4_j\rangle$ only to $|3_k\rangle$ or $|4_k\rangle$, and thus states $|1_j4_k\rangle$ are left invariant, $U_{jk}|1_j4_k\rangle=0=V_{jk}|1_j4_k\rangle$, while $|4_j4_k\rangle$ gain phase due to $V_{jk}|4_j4_k\rangle=V^{44}_{jk}|4_j4_k\rangle$ and $U_{jk}|4_j4_k\rangle=0$. Therefore, the manifold of non-decaying states of $N$ interacting atoms is a $2^N$-dimensional DFS of atoms either in $|1\rangle$ or $|4\rangle$. In this work we consider the case in which initially only a single atom is found in state $|4\rangle$, which is particularly relevant in the context of recent experiments with Rydberg gases~\cite{tiarks_photon_transistor_2014,gorniaczyk_single-photon_2014,gorniaczyk_enhancement_2016,tresp_single-photon_2016,tiarks_optical_2016,thomson_symmetry_2017}. Here, since the dynamics \eqref{eq:master} conserves the number of atoms in $|4\rangle$,  stationary states form a $N$-dimensional DFS, spanned by localised excitations $|\mathbf{j}\rangle := |1_1\cdots 1_{j-1} \,4_j\,1_{j+1}\cdots 1_N\rangle$, or equivalently SWs $|\Psi(\mathbf{k}_s)\rangle=N^{-1/2}\sum_{j=1}^N e^{i\mathbf{k}_s\cdot \mathbf{r}_j}|\mathbf{j}\rangle$, as shown in Fig.~\ref{fig:1}c. Once the weak probe field is switched on, this DFS becomes metastable~\cite{macieszczak_metastability_2016}. On the level of the master operator $\mathcal{L}$ [Eq.~\eqref{eq:master}] this means that the first $N^2$ eigenmodes no longer have strictly zero eigenvalue, but still are separated from the rest of rapidly decaying modes, as sketched in Fig.~\ref{fig:1}b. The dynamics within this metastable DFS, represented by the low-lying modes, then appears stationary at timescales much longer than the relaxation time $\tau_{0,\mathrm{int}}$ of the fast modes, which coincides with the system relaxation in the absence of the probe field, cf. the non-interacting case above. 
\begin{figure}[ht!]
\begin{center}
\includegraphics[width=\columnwidth]{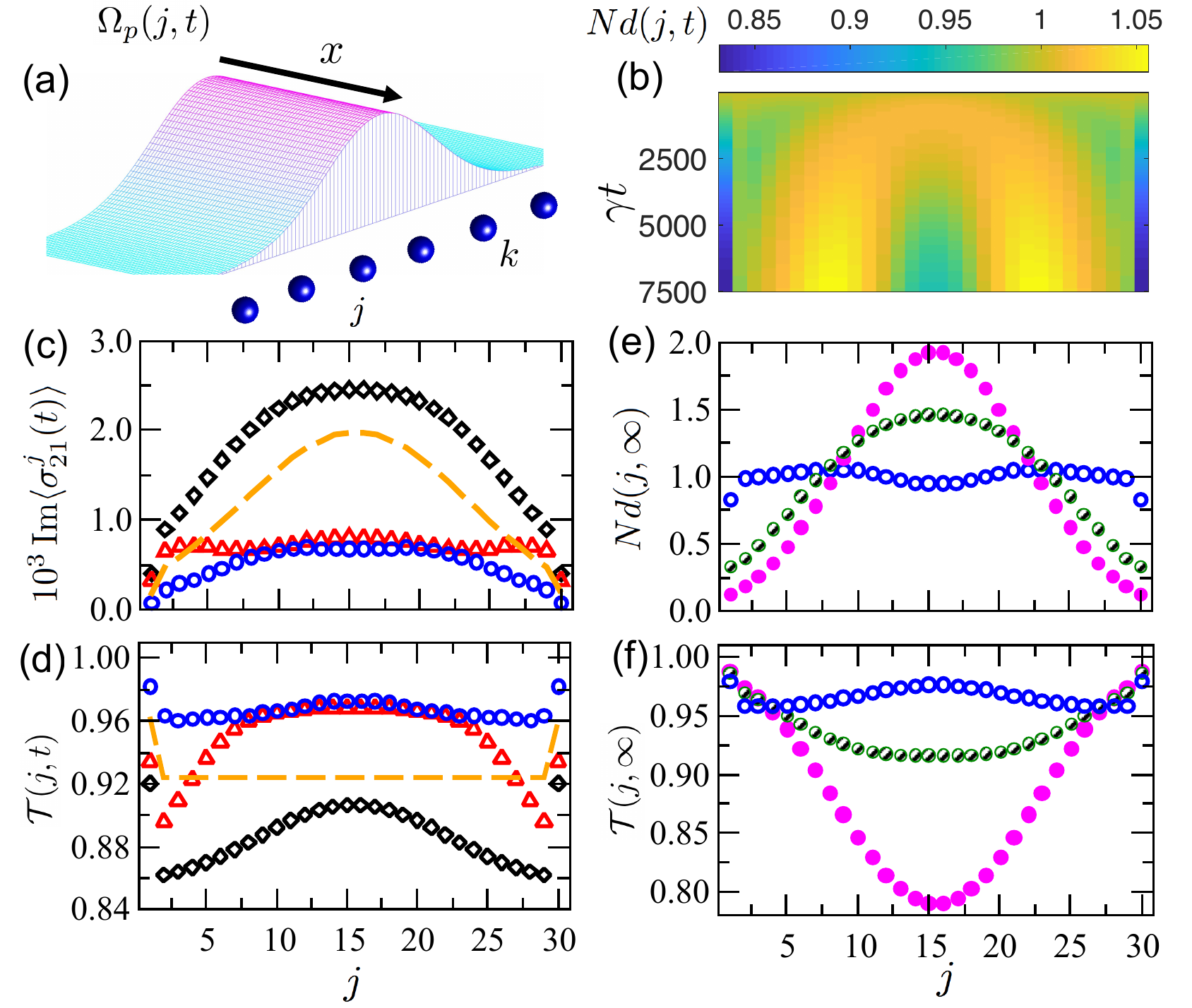}
\vspace*{-6mm}
\caption{
{\bf Long-time dynamics and optical response:}
(a) Stationary Gaussian probe field $\Omega_p(j)$ propagating perpendicular to a chain of $N=30$ atoms with van der Waals interactions, initially in a SW with $\mathbf{k}_s=0$.
(b) Dynamics of the density $d(j,t)$ of a single $|4\rangle$-excitation. The interplay between coherent and dissipative hoping leads to a double-peak density distribution,  strikingly distinct from the uniform density of the non-interacting case. (c) The stationary density $d(j,\infty)$ for lattice spacing $a'=a$ (${\color{blue}\Circle}$), $a'=0.9a$ (${\color{dgreen}\bullet}$), $a'=0.8a$ (${\color{hotmagenta}\CIRCLE}$). In the limit of strong interaction the stationary density follows
the probe field intensity profile as $\rho_{ss}\propto\sum_{j=1}^N|\Omega_p(\mathbf{r}_j)|^2|\mathbf{j}\rangle\!\langle\mathbf{j}|$. (d) The polarisation $\mathrm{Im}\langle \sigma_{21}^j(t)\rangle$ at times $t=100\gamma^{-1}$ ({$\Diamond$}),  $t=1000\gamma^{-1}$ (${\color{red}\vartriangle}$) and $t=7500\gamma^{-1}$ (${\color{blue}\Circle}$). At all times polarisation is distinct from a Gaussian profile that would be observed in the non-interacting case (orange dashed line for $U_{jk}=0$). (e) Exchange interactions lead to a non-uniform transmission profile of the probe field transmission $\mathcal{T}(j,t)$ [cf.~(d) for labels].  (f) The stationary transmission $\mathcal{T}(j,\infty)$ [cf.~(c) for labels] becomes dependent only on the field intensity for strong interactions, see Eq.~\eqref{eq:polINT}.
The probe field profile is  $\Omega_{p}(j)=\Omega_0\exp[-a'^2(j-j_c)^2/2l^2]$ with its center $j_c a'$ coinciding with the chain center,  width $l=8a'$ and $\Omega_0=\Omega_c/20=\gamma/20$. Other parameters as in Fig.~\ref{fig:1}b.
}\vspace*{-8mm}
\label{fig:2}
\end{center}
\end{figure}
%
\section{Effective equations of motion~\label{sec:Leff}} In the following we characterize the dynamics within the metastable DFS, which we calculate analytically to leading order in the probe field strength~\cite{zanardi_dissipative_2016,macieszczak_metastability_2016}, see  Appendices~\ref{app:Icomparison1} and~\ref{app:Iderivation}. For the sake of simplicity we consider here a control field of the form $\Omega_c(\mathbf{r})=e^{i\mathbf{k}_c\cdot\mathbf{r}}|\Omega_c|$ and isotropic density-density interactions, $V_{jk}^{34}=V_{kj}^{34}$. Within the metastable DFS the density matrix $\rho$ evolves according to the master equation
\begin{eqnarray}
\frac{\mathrm{d}}{\mathrm{d}t}\rho=\sum_{j=1}^N \left( -i\left[\sum_{k>j}^N H_{jk},\rho\right]+ \mathcal{D}(L_{j}) \rho\right).\label{eq:Leff}
\end{eqnarray} 
The perturbative Hamiltonians are given by
\begin{eqnarray}
 H_{jk}&=&\omega_{jk}^{z} \,|\mathbf{j}\rangle\!\langle \mathbf{j}|+\omega_{kj}^{z} \,|\mathbf{k}\rangle\!\langle \mathbf{k}|+\left(\omega_{jk}^{xy} \,|\mathbf{j}\rangle\!\langle \mathbf{k} |+\text{h.c.}\right), \label{eq:HamZXY} \end{eqnarray}
with the frequencies
\begin{eqnarray}
\omega_{jk}^{z} &=& \lvert\Omega_p(\mathbf{r}_k,t)\rvert^2\,\mathrm{Im}(\alpha_{jk})\label{eq:omegaZ},\\
\omega_{jk}^{xy}&=& e^{i\mathbf{k}_c\cdot(\mathbf{r}_{j}-\mathbf{r}_{k})}\,\Omega_p(\mathbf{r}_j,t)\,\Omega_p^*(\mathbf{r}_k,t)\,\frac{\beta_{jk}-\beta_{kj}^*}{2i}\label{eq:omegaX},
\end{eqnarray}
where 
\begin{eqnarray}
\alpha_{jk}&=& i \frac{W_{jk}\,|\Omega_c|^2+\eta\, (W_{jk}^2-|U_{jk}^{34}|^2)}{(|\Omega_c|^2+\eta W_{jk})^2-\eta^2|U_{jk}^{34}|^2},\label{eq:alpha}\\
\beta_{jk}&=& i\frac{U_{jk}^{34}\,|\Omega_c|^2}{(|\Omega_c|^2+\eta W_{jk})^2-\eta^2|U_{jk}^{34}|^2},\label{eq:beta}
\end{eqnarray}
with $\eta:= -\delta_2+i\frac{\gamma}{2}$ and $W_{jk}:=\delta_3+V^{34}_{jk}$.  These parameters also enter the jump operators,
\begin{eqnarray}
L_{j}&=&\sqrt{\gamma} \,\sum_{k\neq j}^N\Big(e^{i\mathbf{k}_c\cdot\mathbf{r}_{j}}\,\Omega_p(\mathbf{r}_j,t)\,\alpha_{kj} \,|\mathbf{k}\rangle\!\langle \mathbf{k} |\,+\nonumber\\
&&\quad\,\,+\,e^{i\mathbf{k}_c\cdot\mathbf{r}_{k}}\,\Omega_p(\mathbf{r}_k,t)\,\beta_{kj}\,|\mathbf{k}\rangle\!\langle \mathbf{j} |\Big), \label{eq:jump}
\end{eqnarray}
which correspond to a dissipative decay of an ($j$-th) atom to its ground state after a low-energy excitation is introduced to the system by the probe field. Note that effective dissipative processes within the DFS are in general dependent on coherences between distant sites.\\

When the exchange interaction is zero, $U_{jk}= 0$, we have  $\beta_{jk}= 0$ and the jump operators~\eqref{eq:jump} lead to dephasing between localised excitations, similarly as in the non-interacting case, but with the rates modified by density-density interactions. In contrast, a finite exchange interaction introduces non-local dynamics, through both coherent and dissipative processes. To illustrate this, we study the evolution of the local density $d(j,t)=\langle\mathbf{j}|\rho(t)|\mathbf{j}\rangle$, of atoms in state $|4\rangle$ under the action of a probe field propagating in the direction perpendicular to an atom chain ($z$-axis), as shown in Fig.~\ref{fig:2}a. The field has a stationary Gaussian profile. 
The exchange interaction leads to spatially dependent dynamics of excitations, as the non-uniform field breaks the translation symmetry. As a consequence, for moderate interaction strengths the excitation density dynamically develops a double peak structure from an initially uniform distribution when $V^{34}_{jk}+\delta_3>U_{jk}^{34}$ (Fig.~\ref{fig:2}b), while for strong interactions classical detailed balance dynamics emerges, see Appendix~\ref{app:Istrong}, leading to the stationary state approximately following the probe field profile, $\rho_{ss}\approx \mathcal{N}^{-1} \sum_{j=1}^N|\Omega_p(\mathbf{r}_j)|^2|\mathbf{j}\rangle\!\langle\mathbf{j}|$, where $\mathcal{N}=\sum_{j=1}^N|\Omega_p(\mathbf{r}_j)|^2$,  see Fig.~\ref{fig:2}c. 

\section{Optical response~\label{sec:pol}}  
Let us now study the optical response. For an initial state $\rho$ lying within the single-excitation DFS, the optical response is determined by the polarization
\begin{eqnarray}
&&\langle \sigma_{12}^j(t)\rangle=-\,i \,\Omega_p(\mathbf{r}_j,t)\sum_{k\neq j}^N\alpha_{jk}\, \rho_{kk}+\label{eq:polINT}\\
&& -\,i\sum_{k\neq j}^Ne^{i\mathbf{k}_c\cdot(\mathbf{r}_{k}-\mathbf{r}_{j})}\,\Omega_p(\mathbf{r}_k,t)\,\beta_{jk}\,\rho_{jk} +\mathcal{O}\left(|\Omega_p(\mathbf{r}_k,t)\tau_{0,\mathrm{int}}|^3\right), \nonumber
\end{eqnarray}
where $\rho_{jk}=\langle\mathbf{j}|\rho(t)|\mathbf{k}\rangle$ are coherences between $|4\rangle$- excitations of different atoms, cf.~\cite{li_electromagnetically_2014,coherence_Li_2015}. Compared to the response encountered in conventional EIT \cite{fleischhauer_electromagnetically_2005}, there are two differences. First, there are non-local contributions, i.e. the response of one atom generally depends  on all others. Second, the coherence $\rho_{jk}$ evolves slowly within the metastable manifold, indicating the emergence of a non-equilibrium polarization.

These effects can be seen in Fig.~\ref{fig:2}d, where we show the imaginary part of the polarization and observe a slow change, on a timescale $\propto 1/[|\Omega_p(\mathbf{r}_j)|^{2}\tau_{0,\mathrm{int}}]$~\footnote{As the probe field perturbs a DFS, the long-time dynamics timescales are actually determined by non-dissipative relaxation time $\tau_{0,\mathrm{int}}'\leq\tau_{0,\mathrm{int}}$, given by the inverse of the imaginary gap of the effective Hamiltonian of the dynamics $\mathcal{L}$ at $\Omega_p=0$, i.e. $H-\frac{i}{2}\sum_{j=1}^N\sigma_{22}^{j}$ instead of $\tau_{0,\mathrm{int}}$, cf.~\cite{albert_adiabatic_2016}, see Appendix~\ref{app:Iderivation}.}, from its metastable value to the stationary one. Note, that the timescale corresponding to each atom is not simply monotonically dependent on the probe field $\Omega_p(\mathbf{r}_j)$ due to non-local exchange of the coherence and probe field, cf. Eq.~(\ref{eq:polINT}). In Rydberg experiments, signatures of this physics can be probed through the transmission signal of the probe light as shown in Fig.~\ref{fig:2}e,f. Here we show the transmission  $\mathcal{T}(j,t)=\Delta t^{-1}|\Omega_p(\mathbf{r}_j)|^{-2}\int_{t}^{t + \Delta t} \mathrm{d}t'|\Omega_p^{\text{(out)}}(\mathbf{r}_j,t')|^2$ with $\Delta t =\gamma^{-1}$. In Fig.~\ref{fig:2}e we observe that the signal changes from the initial Gaussian profile to a significantly flatter one at later times. At all times the signal is strikingly different from the uniform and time-independent transmission in the non-interacting case. For stronger interactions the stationary transition simply decreases with the increasing intensity of the probe field, see~Fig.~\ref{fig:2}f.

\section{Stationary states of the interacting system~\label{sec:rhoss}} The stationary state of the long-time dynamics of Eq.~\eqref{eq:Leff} corresponds to the stationary state $\rho_\mathrm{ss}$ of the full dynamics of Eq.~\eqref{eq:master}~\cite{macieszczak_metastability_2016}. Without exchange interactions, the long-time dynamics leads to dephasing of coherences between localised excitations $|\mathbf{j}\rangle$. For an initial state with a single excitation in state $|4\rangle$ there are thus $N$ possible stationary states, with any SW decaying to the fully mixed state, $\rho_\mathrm{ss}=N^{-1}\sum_{j=1}^N| \mathbf{j}\rangle\!\langle\mathbf{j}|+\mathcal{O}(\Omega_p(t)\tau_0)$~\footnote{This structure is not changed by higher order corrections, as all the eigenmodes of the full dynamics in Eq.~\eqref{eq:master} are separable, thus guaranteeing locality of the dynamics}.

To gain some analytic insights into the case of non-zero exchange interactions we consider the cases of all-to-all and nearest-neighbour (NN) interactions. In the former case, $\alpha_{jk}=\alpha$ and $\beta_{jk}=\beta$, which in the presence of the uniform probe field, $\Omega_p(\mathbf{r},t)=e^{i \mathbf{k}_p\cdot\mathbf{r}}|\Omega_p(t)|$, leads to the unique and uniform stationary state,
\begin{eqnarray*}
\rho_\mathrm{ss}&=& N^{-1}\sum_{j=1}^N \Big( | \mathbf{j}\rangle\!\langle\mathbf{j}| +c_N\!\sum_{k\neq j}^N \!e^{i(\mathbf{k}_p+\mathbf{k}_c)\cdot(\mathbf{r}_{j}-\mathbf{r}_{k})}| \mathbf{j}\rangle\!\langle\mathbf{k}|\Big)\!,
\end{eqnarray*}
where $c_N=[(N\!-\!2)|\beta|^2-2\mathrm{Re}(\alpha^*\beta)]/[(N\!-\!1)|\beta|^2+|\alpha|^2]$. For a finite number $N>2$ of atoms, $\rho_\mathrm{ss}$ is mixed unless the resonance $\alpha=-\beta$ takes place. 
\begin{figure}[ht!]
\begin{center}
\includegraphics[width=\columnwidth]{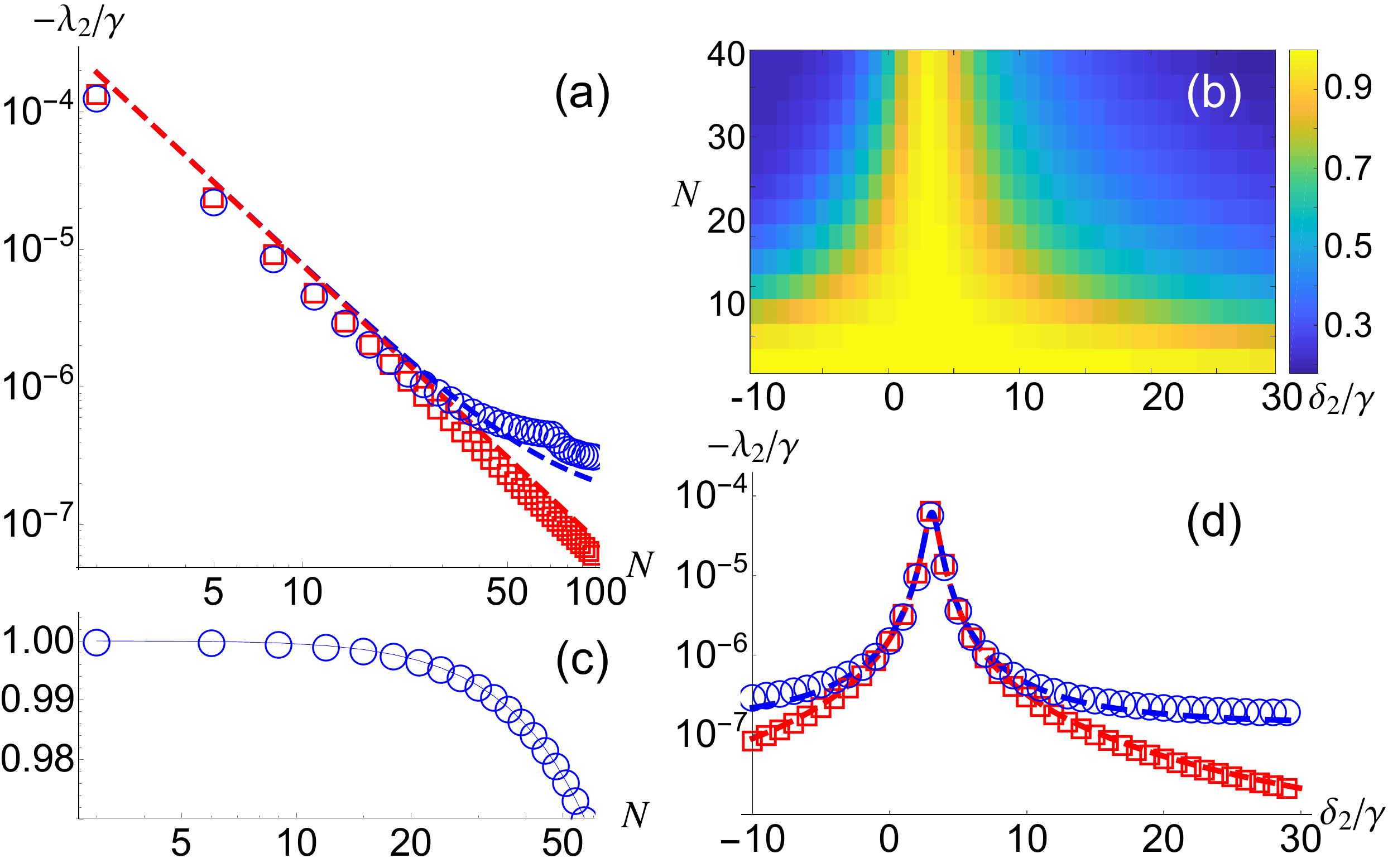}
\vspace*{-4mm}
\caption{
{\bf Spectral gap and pure stationary state}:
(a) Scaling of the spectral gap for an open chain of $N$ atoms with nearest-neighbour (NN) ({\color{red}$\square$}) and van der Waals (vdW) interactions ({\color{blue}$\Circle$}) at $\delta_2=0$. The gap is compared with the scaling $\pi^2\Gamma /N^2$ (red dashed lines) and $\pi^2 \Gamma/N^2+8\Gamma^{(1)}$ (blue dashed lines), where $\Gamma=\gamma|\Omega_p|^2|\alpha_{j,j+1}|^2$ and $\Gamma^{(1)}=\gamma|\Omega_p|^2\,|\alpha_{j,j+2}|^2$, see Appendix~\ref{app:Icomparison2} for discussion. (b) The overlap $\langle\Psi_{ss}|\rho_\text{ss}|\Psi_{ss}\rangle$ of the stationary state  with~\eqref{eq:rhoss_pure} for a system with vdW interactions. The overlap decays with growing $N$ due to the tails in vdW interactions, but  for each $N$ the detuning $\delta_2$ can be chosen to maximise the overlap.  Vertical cut through panel (b) at $\delta_2=3\gamma$  is shown in (c). (d) The spectral gap dependence on the detuning $\delta_2$ for $N=20$ atoms  ({\color{red}$\square$} with NN, {\color{blue}$\Circle$} with vdW interactions). The largest spectral gap, which is well approximated by $\pi^2\Gamma/N^2$ and $\pi^2\Gamma/N^2+8\Gamma^{(1)}$ (red and blue dashed lines), corresponds to the maximal overlap in (b), thus giving the optimal $\delta_2\approx|\Omega_c|^2/[2\, U_{j,j+1}^{34}]$ when $\Gamma\approx |\Omega_p|^2/\gamma$. In these simulations the dispersion coefficients are $C_{34}=C_{\text{ex}}= 1.3\times \gamma a^6$ and the lattice spacing is $\sqrt{2}a$. The fields are uniform $\Omega_{p}=\Omega_c/20=\gamma/20$ and $\delta_3=0$. 
}\vspace*{-6mm}
\label{fig:3}
\end{center}
\end{figure}
In an atom chain with only nearest neighbours interacting, a more general pure stationary state is reached at the interaction resonance $|U^{34}_{j,j+1}|=|V^{34}_{j,j+1}|$, $\delta_3=0$,
\begin{eqnarray}
\lvert\Psi_\mathrm{ss}\rangle=\mathcal{N}^{-1/2}\sum_{j=1}^N (-1)^j\,\widetilde\Omega_p(\mathbf{r}_j) |\mathbf{j}\rangle, \label{eq:rhoss_pure}
\end{eqnarray}
where the normalisation $\mathcal{N}=\sum_{j=1}^N|\Omega_p(\mathbf{r}_j)|^2$ and $\widetilde\Omega_p(\mathbf{r}_j)=e^{i\mathbf{k}_c\cdot\mathbf{r}_j+i\varphi_j}\Omega_p(\mathbf{r}_j)$ with $\varphi_j=-\sum_{k=1}^{j-1}\varphi_{k,k+1}$ determined by the phase of $U^{34}_{j,j+1}V^{34}_{j,j+1}=e^{i\varphi_{j,j+1}}|U^{34}_{j,j+1}||V^{34}_{j,j+1}|$. 
The stationary state is pure [up to $\mathcal{O}(\Omega_p(t)\tau_{0,\mathrm{int}})$] as a collective dark state of the long time dynamics which read, cf.~Eqns.~(\ref{eq:HamZXY}-\ref{eq:jump}),
\begin{eqnarray*}
&&H_{j,j+1}=\frac{|\omega_{j,j+1}|}{2}\,|\boldsymbol{+}_{j}^R\rangle\! \langle \boldsymbol{+}_{j}^R|,\\
&&L_{j}=\sqrt{\Gamma_j^L}|\mathbf{j\!-\!1}\rangle\! \langle \bm{+}^L_{j}|+  \sqrt{\Gamma_j^R}|\mathbf{j\!+\!1}\rangle \langle \bm{+}_{j }^R|,
\end{eqnarray*}
where $|\bm{+}_j^{R,L}\rangle=[\widetilde\Omega_p(\mathbf{r}_{j\pm 1})|\mathbf{j}\rangle+ \widetilde\Omega_p(\mathbf{r}_j)|\mathbf{j\!\pm\!1}\rangle]/|\Omega_p|$, $\sqrt{\Gamma_j^{R,L}}=\sqrt{\gamma}|\Omega_p|\alpha_{j,j\pm 1}$ and $\omega_{j,j+1}=|\Omega_p|^2 \,\mathrm{Im}\alpha_{j,j+1}$, with $|\Omega_p|$ being the maximum amplitude of the probe field  (see Refs.~\cite{diehl_quantum_2008,kraus_preparation_2008} and similar schemes in Refs.~\cite{Rao2013,Rao2014}). It follows that the stationary polarization $\langle \sigma_{12}\rangle_\mathrm{ss}=0$ in the first order, cf.~\eqref{eq:polINT}. For the uniform probe field, numerical results for up to $N=100$ equally spaced atoms suggest that the pure stationary state of a SW is achieved at times $\tau\approx N^2/(\pi^2\, \Gamma)$, where $\Gamma=\gamma|\Omega_p|^2|\alpha_{j,j+1}|^2$, see Fig.~\ref{fig:3}a. For a Rydberg system with van der Waals (vdW) interactions, although the stationary state is in general mixed, it is closely approximated by~\eqref{eq:rhoss_pure} when setting $\delta_2\approx|\Omega_c|^2/[2\, U_{j,j+1}^{34}]$, see Fig.~\ref{fig:3}b,c.  This choice maximes the gap of the system with NN interactions, see~Fig.~\ref{fig:3}d, so that the vdW interactions act as a perturbation of the NN case, cf.~Fig.~\ref{fig:3}a and Appendix~\ref{app:Icomparison2}. Lastly, we note that in the special case of the resonance $U^{34}_{jk}=-V^{34}_{jk}$ (and thus $\alpha_{jk}=-\beta_{jk}$), the stationary state is pure, $\lvert\Psi_\mathrm{ss}\rangle=\mathcal{N}^{-1/2}\sum_{j=1}^N e^{i\mathbf{k}_c\cdot\mathbf{r}_j}\Omega_p(\mathbf{r}_j)|\mathbf{j}\rangle$, for any range of interactions, also including van der Waals interactions, cf.~Eqns.~(\ref{eq:HamZXY}-\ref{eq:jump}).

\section{Unitary operations within the metastable DFS~\label{sec:unitary}} When the detuning $\delta_3$ and interactions are sufficiently small, we have that $\alpha_{jk}\approx i(V_{jk}^{34}+\delta_3)/|\Omega_c|^2$, $\beta_{jk}\approx iU_{jk}^{34}/|\Omega_c|^2$,  and thus the coherent part of the dynamics~\eqref{eq:Leff} is considerably faster than the rate of the two-body dissipation~\eqref{eq:jump}, i.e.~$|\omega_{jk}^z|,|\omega^{xy}_{jk}|\gg \Gamma_{jk}= \gamma\, (\lvert\Omega_p(\mathbf{r}_k,t)\rvert^2+|\Omega_p(\mathbf{r}_j,t)\rvert^2)(|\alpha_{jk}|^2+|\beta_{jk}|^2)$. Actually, even for arbitrary probe and control fields, when the metastable non-interacting DFS is spanned by $| \tilde{\mathbf{j}}\rangle=|\psi_{\mathbf{r}_1}...\psi_{\mathbf{r}_{j-1}}4_j\psi_{\mathbf{r}_{j+1}}...\psi_{\mathbf{r}_N}\rangle$, $j=1,...,N$, cf.~\eqref{eq:dark}, the long-time system dynamics is unitary in the leading order~\cite{zanardi_coherent_2014,zanardi_geometry_2015,albert_adiabatic_2016}, and governed by the  Hamiltonian
\begin{eqnarray}
\tilde{H}_{jk}&=& \tilde{\omega}_{jk}^{z} | \tilde{\mathbf{j}}\rangle\!\langle  \tilde{\mathbf{j}}|+\tilde{\omega}_{kj}^{z}| \tilde{\mathbf{k}}\rangle\!\langle  \tilde{\mathbf{k}}|+\left(\tilde{\omega}_{jk}^{xy} \,| \tilde{\mathbf{j}}\rangle\!\langle  \tilde{\mathbf{k}} |+\text{h.c.}\right)\!,\label{eq:HamXYZ2}%
 \\
\tilde{\omega}_{jk}^{z} &=& |c_{\mathbf{r}_k}|^2 \Big(\delta_3+V^{34}_{kj}+\!\!\!\sum_{l>k,l\neq j}\!\! |c_{\mathbf{r}_l}|^2 V^{33}_{kl} \Big),\nonumber\\
\tilde{\omega}_{jk}^{xy}&=& c_{\mathbf{r}_j}c_{\mathbf{r}_k}^* \,U_{jk}^{34}, \nonumber
\end{eqnarray}
where $c_{\mathbf{r}_j}=\Omega_p(\mathbf{r}_j,t) /\sqrt{|\Omega_p(\mathbf{r}_j,t)|^2+|\Omega_c(\mathbf{r}_j)|^2}$. This can be used to design a fully general unitary evolution in the metastable DFS, assuming it is possible to tune strength of the interactions between pairs of atoms (for dissipative corrections see Appendix~\ref{app:NI}). In such a setup, unitary gates could be performed on the quantum information encoded in collective excitations of SWs~\cite{collective_qubit_07,collective_qubit_08}. 


\section{Summary and conlusions~\label{sec:summary}} We have shown that EIT in an interacting many-body system gives rise to a rather intricate dynamics, featuring a metastable DFS and consequently long timescales. We have derived analytic expressions for the equations of motion in the metastable regime where the open system dynamics feature collective SW dark states. This interesting physics determines the dynamics of both the atomic ensemble and the probe light transmission for example in Rydberg quantum optics experiments and could be probed in detail by Rydberg EIT experiments utilizing two interacting Rydberg states~\cite{tiarks_photon_transistor_2014,gorniaczyk_single-photon_2014,gorniaczyk_enhancement_2016,tresp_single-photon_2016,tiarks_optical_2016,thomson_symmetry_2017,Olmos2011}. Moreover, the dynamics of the atomic ensemble could be applied in the context of all-optical quantum computing, i.e. for the creation of entangled many-body states and the realization of unitary operations on collectively encoded qubits. An interesting future problem concerns the investigation of the coupled collective dynamics of the ensemble and a propagating probe field.

\begin{acknowledgements}
\textit{Acknowledgements.}  K.M. acknowledges discussions with M. M\"uller and D. Viscor. Y.L.Z. acknowledges discussions with A.P. Mandoki. The research leading to these results has received funding from the European Research Council under the European Union's Seventh Framework Programme (FP/2007-2013) / ERC Grant Agreement No.~335266 (ESCQUMA), the EPSRC Grant No. EP/M014266/1, the H2020-FETPROACT-2014 Grant No.~640378 (RYSQ), the German Research Foundation (Emmy-Noether-grant HO 4787/1-1, GiRyd project HO 4787/1-3, SFB/TRR21 project C12), the Ministry of Science, Research and the Arts of Baden-W\"{u}rttemberg (RiSC grant 33-7533.-30-10/37/1), and National Natural Science Foundation of China (Grant No.~11304390 and No.~61632021), and National Basic Research Program of China (Grant No.~2016YFA0301903). 
\end{acknowledgements}

\begin{appendix}



\section{Derivations of long-time dynamics and optical response} \label{app:Iderivation}

Here we derive the long-times dynamics and optical response given in Eqns.~(\ref{eq:Leff}-\ref{eq:polINT}). We use perturbation theory for linear operators~\cite{Kato1995} and consider a weak probe field $\Omega_p(\mathbf{r})$ as a perturbation for dynamics of  $N$ four-level atoms with exchange and density-density interactions in the presence of a uniform control field, i.e. $\frac{\mathrm{d}}{\mathrm{d}t}{\rho}=\L\rho=(\mathcal{L}_0+\mathcal{L}_1)\rho$, where
\begin{eqnarray}
&&\mathcal{L}_0 \rho=
-i\sum_{j=1}^N \bigg[ \delta_2\sigma_{22}^{j}+\delta_3\sigma_{33}^{j}+\left(\Omega_c e^{i\mathbf{k}_c\cdot\mathbf{r}_j}\sigma_{32}^{j}+\text{h.c.}\right)\qquad\label{eq:Imaster0}
\\
&&\quad+\sum_{k>j}^N(U_{jk}+V_{jk}),\rho\,\bigg]+\gamma \sum_{j=1}^N \left(\sigma_{12}^{j}\,\rho\,\sigma_{21}^{j}-\frac{1}{2}\{\sigma_{22}^{j},\rho\}\right)\!,\nonumber\\ 
&&\mathcal{L}_1 \rho=
-i \sum_{j=1}^N \left[\Omega_p(\mathbf{r}_j)\sigma_{21}^{j}+\text{h.c.},\rho\,\right]\!.\label{eq:Imaster1}
\end{eqnarray}

\emph{Long-time dynamics}. As the weak probe field perturbs the stationary DFS of $\L_0$, slow dynamics are induced inside, which can be approximated by the first- and second-
order corrections of the perturbation theory for low-lying eigenmodes of $\L_0+\L_1$~\cite{macieszczak_metastability_2016}. 

\emph{The first-order correction}, $\P_0 \L_1 \P_0$ with $\P_0$ denoting the projection of an initial state on the stationary DFS, corresponds to the unitary dynamics~\cite{zanardi_coherent_2014,zanardi_geometry_2015,macieszczak_metastability_2016}. For~(\ref{eq:Imaster0}-\ref{eq:Imaster1}), we have $\P_0 \L_1 \P_0 \rho=0$, as the weak probe field creates coherences to the outside of the DFS, which decay to $0$ according to the effective Hamiltonian $H_0^\text{eff}$, i.e. from $\P_0=\lim_{t\rightarrow\infty} e^{t\L_0}$, we have $\P_0\L_1\rho=\sum_{j=1}^N\lim_{t\rightarrow\infty} (ie^{-i t H_0^\text{eff}} \sigma^j_{21} \rho +i\rho\sigma^j_{21}e^{i t H_0^{\text{eff}\dagger}})=0$, where $H_0^\text{eff}=\sum_{j=1,k>j}^N H_0^{jk,\text{eff}}$ and  $H_0^{jk,\text{eff}}=V_{jk}+U_{jk}+(\delta_2-i\frac{\gamma}{2}) (\sigma_{22}^{j}+\sigma_{22}^{k})+\delta_3(\sigma_{33}^{j}+\sigma_{33}^{k})+\big[\Omega_c\big(e^{i\mathbf{k}_c\cdot\mathbf{r}_j}\sigma_{32}^{j}+e^{i\mathbf{k}_c\cdot\mathbf{r}_k}\sigma_{32}^{k}\big)+\text{h.c.}\big] $, cf.~\cite{albert_adiabatic_2016}.

\emph{The second-order correction} is $-(\P_0 \L_1 \S_0 \L_1 \P_0 )$ with $\S_0$ being the reduced resolvent of $\L_0$ at $0$, $S_0\L_0=\L_0 S_0=\mathcal{I}-\P_0$~\cite{Kato1995}. It corresponds to completely positive trace-preserving dynamics~\cite{zanardi_coherent_2014,zanardi_geometry_2015,zanardi_dissipative_2016, macieszczak_metastability_2016}. For~(\ref{eq:Imaster0}-\ref{eq:Imaster1})  the perturbation creates coherences to the outside of the DFS, whose decay is described by the effective Hamiltonian $H_0^\text{eff}$. Thus, the resolvent $\S_0=\lim_{t\rightarrow\infty} \int_{0}^t \mathrm{d}t'\,(e^{-i t'\L_0}-\P_0)$ is replaced by the reduced resolvent  $S_0^{\text{eff} \dagger} $ of $H_0^\text{eff}$  at $0$,
\begin{equation}
-(\P_0 \L_1 \S_0 \L_1 \P_0 )\,\rho = - \P_0 \L_1  \sum_{j=1}^N \left( \Omega_p(\mathbf{r}_j)\, S_0^{\text{eff}}\sigma^j_{21} \rho + \text{h.c.}    \right)\label{eq:I2nd}
\end{equation} 
as $\S_0 (-i\sigma^j_{21} \rho)= \lim_{t\rightarrow\infty} \int_{0}^t \mathrm{d}t'\,  (-i)e^{-i t'H_0^\text{eff}} \sigma^j_{21} \rho = S_0^\text{eff}\sigma^j_{21} \rho$, cf.~\cite{albert_adiabatic_2016}. Furthermore, for an initial single excitation to $|4\rangle$, the probe field $\Omega_p(\mathbf{r}_j)$, excites only the eigenmodes of two-body dynamics between the $j$-th atom and the atom excited to $|4\rangle$, $e^{t\L_0}(-i\sigma^j_{21} \rho) = -i e^{-it\sum_{k\neq j}^N H_0^{jk,\text{eff}}}\sigma^j_{21} \rho$, so that $\S_0 (-i\sigma^j_{21} \rho)=\sum_{k\neq j}^N S_0^{jk,\text{eff}}\sigma^j_{21} \rho$, with $S_0^{jk,\text{eff}}$ being the reduced resolvent of $H_0^{jk,\text{eff}}$. Consider now the second perturbation by the probe field, $\Omega_p(\mathbf{r}_l)$ in~\eqref{eq:I2nd}. All atoms except $j$-th, $k$-th and $l$-th are found in the ground level $|1\rangle$ which is disconnected from the dynamics $\L_0$,  and thus the field $\Omega_p(\mathbf{r}_l)$ introduces dynamics of at most $N=3$ atoms, so that $\P_0$ is replaced  in~\eqref{eq:I2nd} by the projection on the DFS of those atoms, $\P_0^{jkl}$. Furthermore, note that for $l\neq j,k$ we have $\sigma_{12}^{l} S_0^{jk,\text{eff}}\sigma^j_{21} \rho=0$, while $\sigma_{21}^{l} S_0^{jk,\text{eff}}\sigma^j_{21} \rho  = S_0^{jk,\text{eff}}\sigma_{21}^{l}\sigma^j_{21} \rho $ (and their conjugations) decay to $0$, i.e. $\P_0(\sigma_{21}^{l} S_0^{jk,\text{eff}}\sigma^j_{21} \rho )= \lim_{t\rightarrow\infty} e^{-i t H_0^\text{eff}}(S_0^{jk,\text{eff}}\sigma_{21}^{l}\sigma^j_{21} \rho)=0$, cf.~\cite{albert_adiabatic_2016}. Therefore, for $l\neq j,k$ only terms $
\P_0( S_0^{jk,\text{eff}}\sigma^j_{21} \rho \sigma_{12}^{l})$ contribute, and 
$\P_0^{jkl}$ can be replaced by the corresponding projection from the subspace featuring only two excitations. Eqns.~(\ref{eq:Leff}-\ref{eq:jump}) of the main text follow.\\

\emph{Optical response.} The metastable states up to second order corrections are given by~\cite{Kato1995,macieszczak_metastability_2016},
\begin{eqnarray}
\rho-\S_0 \L_1 \P_0\rho
&=&\rho -\! \sum_{j=1}^N\! \bigg(\! \Omega_p(\mathbf{r}_j) \sum_{k\neq j}^N S_0^{jk,\text{eff}}\sigma^j_{21} \rho + \text{h.c.} \bigg), \quad \qquad\label{eq:IrhoMS}
\end{eqnarray}
where $\rho$ is in the stationary DFS, and we assumed a single $|4\rangle$-excitation in the system, cf. discussion below~\eqref{eq:I2nd}. In particular, coherence between the levels $|1\rangle$ and $|2\rangle$ is induced, 
\begin{eqnarray}
\langle \sigma_{12}^l\rangle_\rho&\approx& - \sum_{j=1}^N \sum_{k\neq j}^N  \Omega_p(\mathbf{r}_j)\,\mathrm{Tr} (  \sigma_{12}^l\, S_0^{jk,\text{eff}}\sigma^j_{21} \rho  )\nonumber\\\nonumber
&=& - \Omega_p(\mathbf{r}_l)\sum_{k\neq l}^N  \mathrm{Tr} (  \sigma_{12}^l\, S_0^{lk,\text{eff}}\sigma^l_{21} \rho  )+ \\
&&- \sum_{j\neq l}^N  \Omega_p(\mathbf{r}_j)\,\mathrm{Tr} (  \sigma_{12}^l\, S_0^{jl,\text{eff}}\sigma^j_{21} \rho  ),\qquad 
\label{eq:Ipol}
\end{eqnarray}
where the equality follows from the fact that for $S_0^{jk,\text{eff}}\sigma^j_{21} \rho$ all atoms except $j$-th, $k$-th are found in the ground level $|1\rangle$. The local and non-local contributions to the polarization lead directly to Eq.~\eqref{eq:polINT} by solving the  first-order corrections for $N=2$ atoms.

\section{Effective dynamics vs. full dynamics for few atoms}\label{app:Icomparison1}

For  $N=3$ atoms we compare the effective dynamics in the $N$-dimensional DFS, Eqns.~(\ref{eq:Leff}-\ref{eq:jump}), with  the dynamics on the full Hilbert space of a dimension $4^N$. In Tab.~\ref{tab:1} we show the excitation density, $d(j)=\langle\mathbf{j}|\rho|\mathbf{j}\rangle$, and the polarization, $\langle \sigma_{12}^j\rangle$, for the stationary states of the effective dynamics in the DFS, $\rho_\text{DFS}$,  and the exact solution on the full system space, $\rho_\text{full}$, together with fidelity between those states, $F=[\mathrm{Tr}(\rho_\text{DFS}^{1/2}\,\rho_\text{full}\,\rho_\text{DFS}^{1/2})^{1/2}]^2$.   Results of Tab.~\ref{tab:1} agree with predictions of higher order corrections in the probe field: quadratic for the density, cubic for the polarisation, and quadratic for the infidelity $1-F$~\cite{Kato1995}.
\begin{table}[htb!]
\begin{center}
\begin{tabular}{c || c|c}
\textbf{Dynamics}&Effective&Full\\ \hline \hline
$d(1)=d(3)$& $ 0.3515570  $&$0.3512607$\\
$d(2)$&$0.296886$&$0.296615$\\
$[\langle \sigma_{12}^1\rangle\!=\!\langle \sigma_{12}^3\rangle]\!\times\! 10^3$&$ 1.70451 -i\, 0.379854$ &$1.70697 - i\,0.379947 $\\
$\langle \sigma_{12}^2\rangle\times 10^3$&$5.33015 - i\,1.39312$&  $5.33241 - i\,1.39369$\\
\hline
$F$& \multicolumn{2}{c}{$0.9996$}\\
\hline
\end{tabular}
\caption{Stationary excitation density, polarization and fidelity for the stationary states of $N=3$ atoms with vdW interactions, $C_{34}=1.3\times \gamma a^6$, $C_{\text{ex}}= 1.0\times \gamma a^6$, in a lattice with spacing $a$ and open boundaries. The fields are uniform $\Omega_{p}=\Omega_c/50=\gamma/50$, and detunings $\delta_3=\delta_2=0$. \label{tab:1}} \vspace*{-10mm}
\end{center}
\end{table}

\section{Dynamics in the limit of strong interactions} \label{app:Istrong}

In the limit of the strong interactions, $|U_{jk}^{34}|\rightarrow\infty$ or $|V_{jk}^{34}|\rightarrow\infty$, we have
\begin{eqnarray}
\alpha_{jk}&\rightarrow&  \frac{i}{\eta}=\frac{-i\,\delta_2+\gamma/2}{\delta_2^2+\gamma^2/4},\qquad
\beta_{jk}\rightarrow 0, \label{eq:alphabetaI}
\end{eqnarray}
cf.~(\ref{eq:alpha}-\ref{eq:beta}), which is analogous to the non-interacting EIT with large detuning $\delta_3\rightarrow\infty$. The stationary state is $N$-degenerate, $\rho_\text{ss}=\sum_{j=1}^N p_j|\mathbf{j}\rangle\!\langle\mathbf{j}|$, with the probability distribution determined by the initial excitation density, $p_j=\langle\mathbf{j}|\rho_{0} |\mathbf{j}\rangle$. Note that at the interaction resonance, $U^{34}_{jk}/V^{34}_{jk}\rightarrow e^{\phi_{jk}}$, Eq.~\eqref{eq:alphabetaI} is not valid, as $\alpha_{jk}\rightarrow\frac{i}{2\eta}$ and $\beta_{jk}\rightarrow\frac{i }{2\eta}e^{\phi_{jk}}$, which leads to the unique stationary state (see e.g. Eq.~\eqref{eq:rhoss_pure}).	
 
Away from the resonance the degeneracy of a stationary state $\rho_\text{ss}$ is lifted by the non-local corrections to \eqref{eq:alphabetaI} as follows. First, an initial state   dephases into a mixture of localised excitations, and coherence $|\mathbf{j}\rangle\!\langle{\mathbf{k}}|$, $j\neq k$, decays at a rate $\Gamma_j+\Gamma_k$ and oscillation frequency  $\omega_j^z-\omega_k^z$,  where  $\Gamma_j=\gamma\sum_{j'\neq j}|\Omega_p(\mathbf{r}_{j'})|^2|\alpha_{jj'}|^2/2$ and $\omega_j^z=\sum_{j'\neq j}|\Omega_p(\mathbf{r}_{j'})|^2 \mathrm{Im}\,\alpha_{jj'}$.  At later times $t\gg\tau_\alpha=\max_{j,k, j\neq k}[( \Gamma_j+\Gamma_k)^2+(\omega_j^z-\omega_k^z)^2]^{-1/2}$, this is followed by classical evolution, $\rho(t)=\sum_{j=1}^N p_j(t)|\mathbf{j}\rangle\!\langle\mathbf{j}|$, with
\begin{eqnarray}
\frac{\mathrm{d}}{\mathrm{d}t} p_j (t)&=& \sum_{k=1}^N \left[(T_1)_{jk}+(T_2)_{jk}\right] \, p_k(t),\label{eq:T}
\end{eqnarray}
where for $j\neq k$ 
\begin{eqnarray}
(T_1)_{jk}&=& \gamma|\Omega_p(\mathbf{r}_j)|^2 |\beta_{jk}|^2, \label{eq:T12}\\
(T_2)_{jk}&=& |\Omega_p(\mathbf{r}_j)|^2|\Omega_p(\mathbf{r}_k)|^2\,|\beta_{jk}|^2\,\mathrm{Re}\,\frac{2+2\gamma\alpha_{jk}}{\Gamma_j+\Gamma_k+ i(\omega_j^z-\omega_k^z) },
\nonumber
\end{eqnarray}
and $(T_{1,2})_{jj}=-\sum_{k\neq j} (T_{1,2})_{kj}$~\cite{Kato1995}. The first contribution $T_1$ is due to exchange-interactions terms in jumps $L_j$, see Eq.~\eqref{eq:jump} and obeys the detailed balance condition, so that its stationary state follows the probe field intensity profile, $\rho_{ss}= \mathcal{N}^{-1} \sum_{j=1}^N|\Omega_p(\mathbf{r}_j)|^2|\mathbf{j}\rangle\!\langle\mathbf{j}|$, where $\mathcal{N}=\sum_{j=1}^N|\Omega_p(\mathbf{r}_j)|^2$.  The second contribution $T_2$ represents density fluctuations due to coherences created between the localised excitations. 
For strong interactions $T_1$ dominates and the stationary state approximately follows the field intensity profile, cf. Fig.~\ref{fig:2}c. We have assumed that the classical dynamics,  Eqns.~(\ref{eq:T}-\ref{eq:T12}), dominate higher-order corrections in the probe field neglected in Eq.~\eqref{eq:Leff}, which is true for a weak enough probe field.  

Derivation of Eqns.~(\ref{eq:T}-\ref{eq:T12}) is based solely on the fact that $|\alpha_{jk}|\gg|\beta_{jk}|$ and $\gamma|\alpha_{jk}|^2\gg|\beta_{jk}|$. Therefore, classical dynamics of~(\ref{eq:T}-\ref{eq:T12}) also arise when the density-density interaction or $\delta_3$-detuning, although finite, dominate the exchange interaction, $|U_{jk}^{34}|\ll |V_{jk}^{34}+\delta_3|$ (unless the interactions are weak and the unitary motion cannot be neglected). 

\section{Timescale of relaxation to  pure stationary state} \label{app:Icomparison2}

For nearest-neighbour (NN) interactions at the resonance, $|V^{34}_{j,j+1}|=|U^{34}_{j,j+1}|$ and $\delta_3=0$, the stationary state of the long-time dynamics is pure, see~\eqref{eq:rhoss_pure}. In Fig.~\ref{fig:3}a the spectral gap for chains of up to $N=100$ equally spaced atoms in the uniform probe field, follows the scaling $(-\mathrm{Re}\lambda_2)\approx \pi^2\Gamma/N^2$, where the nearest-neighbour dissipation rate $\Gamma=\gamma|\Omega_p|^2|\alpha_{j,j+1}|^2$. Below we show analytically that the spectral gap is asymptotically bounded,
\begin{equation}
(-\mathrm{Re}\lambda_2)_\text{p.b.c.}\leq 4\pi^2\Gamma/N^2 \label{eq:gapPBC}
\end{equation} 
for chains with periodic  boundary conditions, and 
\begin{equation}
 (-\mathrm{Re}\lambda_2)_\text{o.b.c.}\leq 12\Gamma/N^2,  \label{eq:gapOBC}
\end{equation} 
for open boundary conditions.

Dynamics of coherences to the dark state, $|\Psi\rangle\!\langle\Psi_\text{ss}|$, are governed by the effective Hamiltonian, $-i H^\text{eff}=\sum_{j=1}^N \big(-i \sum_{k>j}^N H_{jk}-\frac{1}{2} L_j^\dagger L_j \big)$, i.e. $\frac{\mathrm{d}}{\mathrm{d}t}|\Psi\rangle\!\langle\Psi_\text{ss}|=-i H^\text{eff}$. For NN interactions and the uniform probe field, $-i H^\text{eff}=-|\Omega_p|^2\sum_{j} \big[\alpha_{j,j+1}\big(|\mathbf{j}\rangle\!\langle \mathbf{j}|+|\mathbf{j+1}\rangle\!\langle \mathbf{j+1}|\big)+\beta_{j,j+1}\big(e^{i(\mathbf{k}_c+\mathbf{k}_p)\cdot(\mathbf{r}_{j}-\mathbf{r}_{j+1})}|\mathbf{j}\rangle\!\langle \mathbf{j\!+\!1}|+\text{h.c.}\big)\big]$. Considering equally spaced atoms at the interaction resonance $\beta_{j,j+1}=\alpha_{j,j+1}=\alpha$, further gives  $-iH^\text{eff}=-\alpha|\Omega_p|^2\sum_{j=1}^{N-1} |\bm{+}_j\rangle\langle\bm{+}_j|$, where $|\bm{+}_j\rangle=e^{i(\mathbf{k}_c+\mathbf{k}_p)\cdot\mathbf{r}_{j}}|\mathbf{j}\rangle+e^{i(\mathbf{k}_c+\mathbf{k}_p)\cdot\mathbf{r}_{j+1}}|\mathbf{j\!+\!1}\rangle.$

For p.b.c. and an even number $N$ of atoms,  the spin waves $|\Psi(\mathbf{k}_s)\rangle=\frac{1}{N}\sum_{j=1}^N  e^{i\mathbf{k}_s\cdot\mathbf{r}_j} |\mathbf{j}\rangle$, with $\mathbf{k_s}=\mathbf{k}_c+\mathbf{k}_p+ k \frac{2\pi}{ N  a }\mathbf{n}$ and $a\,\mathbf{n}=\mathbf{r}_{j+1}-\mathbf{r}_j $, $k=1,..,N$, are the eigenmodes of $-i H^\text{eff}$ with the eigenvalues $-2\alpha\big[1+\cos(\frac{2\pi k}{ N})\big]$. The choice $k=\frac{N}{2}\pm1$ leads to~\eqref{eq:gapPBC} by noting that $\mathrm{Re}(\alpha)=\gamma|\alpha|^2$.

For a system with o.b.c.,  we use a variational principle for Hermitian $i\alpha^{-1}H^\text{eff}$ in order to find its second eigenvalue above the known minimum, which  equals $0$ and corresponds to $|\Psi_\text{ss}\rangle=\frac{1}{N}\sum_{j=1}^N (-1)^j e^{i(\mathbf{k}_c+\mathbf{k}_p)\cdot\mathbf{r}_j}|\mathbf{j}\rangle$, cf.~\eqref{eq:rhoss_pure}. For the variational set reduced to the spin waves we then have
\begin{eqnarray*}
 \min_{|\Psi\rangle} \frac{\alpha^{-1} \langle\Psi|i H^\text{eff}|\Psi\rangle}{|\langle\Psi|\Psi\rangle|^2-|\langle\Psi_\text{ss}|\Psi\rangle|^2} &&\leq |\Omega_p|^2 \min_{k }  \frac{2\frac{N-1}{N}\big[1+\cos(\frac{2\pi k}{ N})\big]
}{1-\left\lvert\frac{1}{N}\frac{1-e^{i(2k-N)\pi}}{1+e^{i\frac{2 \pi k}{ N}}} \right\rvert^2}\\&&\underset{N\rightarrow\infty}{\approx}\frac{12\, |\Omega_p|^2}{N^2},
\end{eqnarray*}
where the last approximation follows from $k=\frac{N}{2}+x$ and $x\rightarrow 0$ and gives~\eqref{eq:gapOBC}.\\

\emph{Approximately pure state preparation with van der Waals interactions}. In Fig.~\ref{fig:3}a we also show the scaling with $N$ of the gap for the system van der Waals (vdW) interactions, which features a characteristic slowing down absent for NN interactions. For moderate $N$ this is a consequence of van der Waals interactions being a weak next-nearest-neighbour (NNN) perturbation to NN interactions, as $U^{34}_{jk}=C_\text{ex}/|\mathbf{r}_j-\mathbf{r}_k|^{6}$, $V^{34}_{jk}=C_{34}/|\mathbf{r}_j-\mathbf{r}_k|^{6}$. For $C_\text{ex}=C_{34}$, the NNN perturbation is at the opposite resonance to the fulfilled by $|\Psi_\text{ss}\rangle$. For p.b.c. and even $N$ this leads to the eigenvalue shift by $ \frac{8}{N}\sqrt{\Gamma^{(1)}\Gamma}[1+\cos(\frac{4\pi k}{ N})]  + \frac{8}{N}\Gamma^{(1)} [1+\cos(\frac{4\pi k}{ N})] - 2\alpha^{(1)}|\Omega_p|^2[1+\cos(\frac{4\pi k}{ N})]-4[\alpha^{(1)}]^*|\Omega_p|^2 $, where $\alpha^{(1)}=\alpha_{j,j+2}$ and $k\neq N/2$~\cite{Kato1995}. Therefore, for moderate $N$ the bound is modified as
\begin{equation}
(-\mathrm{Re}\lambda_2)_\text{p.b.c.}\leq 4\pi^2\Gamma/N^2+8\Gamma^{(1)}.\label{eq:gapvdW}
\end{equation} 
In Fig.~\ref{fig:3}a,d such a shift describes well the gap scaling also for o.b.c. The influence of NNN interactions can be minimised by the choice of detuning $\delta_2=|\Omega_c|^2/(2\, U_{j,j+1})$, which corresponds to the maximal gap of the system with NN interactions and $\Gamma=\Omega_p^2/\gamma$, see Fig.~\ref{fig:3}b,d. In this case the the stationary state $\rho_{ss}$, although mixed, is close to the pure state of Eq.~\eqref{eq:rhoss_pure}, see Fig.~\ref{fig:3}b,c.

For the van der Waals interactions at the opposite resonance,  $C_\text{ex}=-C_{34}$, the stationary state is pure, $|\Psi_\text{ss}\rangle=\frac{1}{N}\sum_{j=1}^N e^{i(\mathbf{k}_c+\mathbf{k}_p)\cdot\mathbf{r}_j}|\mathbf{j}\rangle$, for o.b.c. the gap scales at least as fast as $12\,(\Gamma+4\Gamma^{(1)})/N^{2}$, since $-\langle\Psi(\mathbf{k}_s)|i H^\text{eff}|\Psi(\mathbf{k}_s)\rangle=-2\alpha\big[1-\cos(\frac{2 \pi k}{ N})\big]\frac{N-1}{N}-2\alpha^{(1)}\big[1-\cos(\frac{4 \pi k}{N})\big]\frac{N-2}{N}$+(...). For p.b.c. we arrive at $4\pi^2(\Gamma+4\Gamma^{(1)})/N^{2}$.

\section{Derivation of unitary dynamics in the limit of small interactions and dissipative corrections} \label{app:NI}
Consider non-interacting dynamics of $N$ 4-level atoms,
\begin{eqnarray}
\mathcal{L}_0\rho&=& \sum_{j=1}^N\! \Big(\!-\!i\left[\delta_2\sigma_{22}^{j}
+\left(\Omega_p(\mathbf{r}_j)\sigma_{21}^{j}+\Omega_c(\mathbf{r}_j)\sigma_{32}^{j}
+\text{h.c.}\right)\!,\rho\right]\nonumber\\
&&\qquad+\,\gamma\,\sigma_{12}^{j}\,\rho\,\sigma_{21}^{j}-\frac{\gamma}{2}\{\sigma_{22}^{j},\rho\}\Big),\label{eq:master2}
\end{eqnarray}
perturbed by small detuning $\delta_3$ and weak density-density, $V_{jk}$, and exchange interactions, $U_{jk}$,
\begin{eqnarray}
\mathcal{L}_1 \rho&=&
-i\sum_{j=1}^N \bigg[ \delta_3\sigma_{33}^{j}+\sum_{k> j}^N \Big( U_{jk}^{34}\,\sigma_{43}^{j}\, \sigma_{34}^{k}+\mathrm{h.c.}\nonumber
\\&&\qquad \qquad \qquad \quad+\!\!\!\!\sum_{n,m=3,4}\!\!\!V_{jk}^{nm}\, \hat{\sigma}_{nn}^{j}\hat{\sigma}_{mm}^{k}\Big),\rho\bigg].\quad
\end{eqnarray}
The stationary DFS of non-interacting $\L_0$ is a tensor product of 2-dimensional DFS of individual atoms spanned by dark $|\psi_\mathbf{r}\rangle=(\Omega_c^*(\mathbf{r})|1\rangle-\Omega_p(\mathbf{r})|3\rangle)/\sqrt{\lvert\Omega_c(\mathbf{r})\rvert^2+\lvert\Omega_p(\mathbf{r})\rvert^2}$ and disconnected $|4\rangle$. Let $\P_0$ denote the projection of an initial state onto the stationary DFS.

\emph{Unitary dynamics} inside the stationary DFS of $\L_0$  are governed by the first-order correction, $ \P_0 \L_1\P_0 $~\cite{Kato1995,zanardi_coherent_2014,zanardi_geometry_2015,macieszczak_metastability_2016}. 
As the perturbation $\L_1$ creates coherences to a dark DFS, whose dynamics is described by the effective Hamiltonian, we have $\P_0 (-i (U_{jk}+V_{jk})\rho) = -i\lim_{t\rightarrow\infty}e^{-it (H_0^\text{j,eff}+H_0^\text{k,eff})}  (U_{jk}+V_{jk})\rho= P_0^{j}\otimes P_0^{k}  (U_{jk}+V_{jk})\rho$, where $H_0^\text{j,eff}=(\delta_2-i\frac{\gamma}{2})\sigma_{22}^{j}
+\big(\Omega_p(\mathbf{r}_j)\sigma_{21}^{j}+\Omega_c(\mathbf{r}_j)\sigma_{32}^{j}
+\text{h.c.}\big)$ and $P_0^{j}$ is the orthogonal projection on the $j$-th atom DFS, cf.~\cite{albert_adiabatic_2016} and Appendix~\ref{app:Iderivation}. Therefore,
\begin{eqnarray}
 &&(\P_0 \L_1\P_0 )\,\rho =   -i\bigg[\sum_{j=1}^N \bigg(|c_{\mathbf{r}_j}|^2 \delta_3  \,|\psi_{\mathbf{r}_j}\rangle\!\langle\psi_{\mathbf{r}_j}|+ \label{eq:NI1storder}\\
 &&  +\!\!\sum_{k=1,k\neq j}^N\! \Big(    U^{34}_{jk}\, c_{\mathbf{r}_k}^*c_{\mathbf{r}_j}\, |4_j \psi_{\mathbf{r}_k}\rangle\!\langle\psi_{\mathbf{r}_j} 4_k |+V^{44}_{jk}\,|4_j 4_k\rangle\!\langle 4_j 4_k|+  \nonumber\\
 && 
+\, |c_{\mathbf{r}_j}|^2  \big( 
 V^{34}_{jk}\,|\psi_{\mathbf{r}_j} 4_k\rangle\!\langle\psi_{\mathbf{r}_j} 4_k|+ V^{33}_{jk}\, |c_{\mathbf{r}_k}|^2    |\psi_{\mathbf{r}_j}\psi_{\mathbf{r}_k}\rangle\!\langle\psi_{\mathbf{r}_j}\psi_{\mathbf{r}_k}| \big)\Big)\bigg),\rho\,\bigg],\nonumber
\end{eqnarray}
where  $c_{\mathbf{r}_j}:=\langle 3|\psi_{\mathbf{r}_j}\rangle=-\Omega_p(\mathbf{r}_j) /(|\Omega_p(\mathbf{r}_j)|^2+|\Omega_c(\mathbf{r}_j)|^2)^{1/2}$ and $U^{34}_{kj}=(U_{jk}^{34})^*$. Eq.~\eqref{eq:NI1storder}  for the case of an initial state with a single $|4\rangle$-excitation gives Eq.~\eqref{eq:HamXYZ2}. 
\\

\emph{Dissipative corrections} are given by~\cite{zanardi_coherent_2014,zanardi_geometry_2015,macieszczak_metastability_2016}
\begin{eqnarray}
\rho(t)&=&e^{t\L}\rho_\text{in}\approx e^{t\P_0 \L_1\P_0}\,\rho_\text{in}+\mathcal{O}(t\tilde{\L}),\label{eq:NIcorr} 
\end{eqnarray}
where the generator of the second-order dynamics 
\begin{eqnarray}
&&\tilde{\L}\, \rho=\sum_{j=1}^N\!\bigg(\!\!-i\left[\tilde{H}_j  ,\rho\right]\!+ \mathcal{D}(\tilde{L}_{j})\rho+\sum_{k> j}^N\mathcal{D}(\tilde{L}_{jk})\rho\! \bigg),\quad\label{eq:NILeff}
\end{eqnarray}
with
\begin{eqnarray}
&&\tilde{L}_{j}=\sqrt{\gamma} \,\frac{\Omega_c(\mathbf{r}_j)}{|\Omega_p(\mathbf{r}_j)|^2+|\Omega_c(\mathbf{r}_j)|^2}\bigg(\!c_{\mathbf{r}_j} \delta_3|\psi_{\mathbf{r}_j}\rangle\!\langle\psi_{\mathbf{r}_j}|+ \label{eq:NIjump1}\\\nonumber
&&+\sum_{k\neq j}^N \big(c_{\mathbf{r}_j} V_{jk}^{34}\, |\psi_{\mathbf{r}_j}4_k\rangle\!\langle\psi_{\mathbf{r}_j}4_k|+ c_{\mathbf{r}_k} U_{kj}^{34} \, |\psi_{\mathbf{r}_j}4_k\rangle\!\langle 4_j \psi_{\mathbf{r}_k}|\big)\! \bigg),  \\
&&\tilde{L}_{j k}=  \sqrt{2\,\mathrm{Re}(s_{jk})}\,c_{\mathbf{r}_j}\,c_{\mathbf{r}_k} V_{jk}^{33}\,|\psi_{\mathbf{r}_j}\psi_{\mathbf{r}_k}\rangle\!\langle\psi_{\mathbf{r}_j}\psi_{\mathbf{r}_k}|,  \label{eq:NIjump2}\\
&&\tilde{H}_{j}=\frac{\delta_2}{\gamma}\, \tilde{L}_j^\dagger\tilde{L}_j-\sum_{k> j}^N \frac{\mathrm{Im}(s_{jk})}{2\,\mathrm{Re}(s_{jk})}\, \tilde{L}_{jk}^\dagger\tilde{L}_{jk}. \label{eq:NIHam}
\end{eqnarray}
The parameter $s_{jk}=-i\,\langle 3_j 3_k|S_0^{jk,\text{eff}}| 3_j 3_k\rangle$, where $S_0^{jk,\text{eff}}$ is the resolvent of $H_0^{j,\text{eff}}+H_{0}^{k,\text{eff}}$. When the control and probe fields are uniform, $s_{jk}= |\Omega_c|^4  \left[(\gamma/2 + i\delta_2)^2+ |\Omega_p|^2+ |\Omega_c|^2 \right]/[2 (\gamma/2 + i\delta_2) \left(|\Omega_p|^2 +|\Omega_c|^2 \right)^3]$. The jump operators $\tilde{L}_{j}$ correspond to a dissipative decay of a single $j$-th atom, while $\tilde{L}_{j k}$ to a coincident decay of $j$-th and $k$-th atom.

\emph{Derivation}. For initial state $\rho$ inside the DFS,  the dynamics is approximated in the second order by~\cite{Kato1995,macieszczak_metastability_2016},
\begin{eqnarray}
&&\tilde\L\,\rho= -(\P_0 \L_1 \S_0 \L_1 \P_0) \rho\label{eq:NI2ndorder0}\\\nonumber
&&=- \P_0 \L_1 \! \sum_{j=1}^N\! \bigg(\! \delta_3S_0^{j,\text{eff}}\sigma^j_{33} \rho +\!\sum_{k>j}^N\! S_0^{jk,\text{eff}} (U_{jk}\!+\!V_{jk}) \rho    +\text{h.c.}  \! \bigg),
\end{eqnarray}
where $\S_0$ is the reduced resolvent for $\L_0$ at 0, and the last equality follows the fact that  the perturbation $\L_1$ creates coherences to the DFS evolving  with the effective Hamiltonian, so that $\S_0 (-i (U_{jk}+V_{jk})\rho)= -i \lim_{t\rightarrow\infty} \int_{0}^t \mathrm{d}t'\,(e^{-i t'(H_0^{j,\text{eff}}+H_0^{k,\text{eff}})}-P_0^j\otimes P_0^k)  (U_{jk}+V_{jk})\rho = S_0^{jk,\text{eff}} (U_{jk}+V_{jk})\rho $. $S_0^{j,\text{eff}}$, $S_0^{jk,\text{eff}}$ are the reduced resolvents at $0$ for $H_0^{j,\text{eff}}$ and $ H_0^{j,\text{eff}}+H_0^{k,\text{eff}}$, respectively. Furthermore, the interactions, $U_{jk}+V_{jk}$, perturb only the dark state $|\psi_\mathbf{r}\rangle$ outside the DFS, but not $|4\rangle$, so that 
\begin{eqnarray*}
S_0^{jk,\text{eff}} U_{jk}\,\rho&=& U_{jk}^{34}\,  \sigma_{43}^j\,(S_0^{k,\text{eff}} \sigma_{34}^k)\, \rho  +U_{kj}^{34}\, (S_0^{j,\text{eff}} \sigma_{34}^j)\, \sigma_{43}^k\, \rho, \label{eq:obs1}\\
S_0^{jk,\text{eff}} V_{jk}\,\rho&=& V_{jk}^{34} \,(S_0^{j,\text{eff}} \sigma_{33}^j)\, \sigma_{44}^k\, \rho  +V_{kj}^{34}\,   \sigma_{44}^j\,(S_0^{k,\text{eff}}\sigma_{33}^k)\,\rho +\\
&&+ V_{jk}^{33}\,S_0^{jk,\text{eff}}\, (\sigma_{33}^j\, \sigma_{33}^k)\, \rho.  \label{eq:obs2}
\end{eqnarray*} 
The only non-zero contributions in~\eqref{eq:NI2ndorder0} come from the second $\L_1$ acting again on the atom perturbed outside the DFS (see below), i.e. $j$-th atom in $U_{kj}^{34}$ and $V_{jk}^{34}$ terms,  or $k$-th atom in $U_{jk}^{34}$ and $V_{kj}^{34}$ terms, or both atoms in $V_{jk}^{33}$ term. Thus, Eqns.~(\ref{eq:NILeff}-\ref{eq:NIHam}) follow from the solution for $N=3$ atoms. 

When the second $\L_1$ does acting on the atom inside DFS, there are terms of two types (and their conjugates), e.g. $X_{lm} \,(S_0^{j,\text{eff}} \sigma_{33}^j)\, \sigma_{44}^k\, \rho = (S_0^{j,\text{eff}} \sigma_{33}^j)\,X_{lm}\, \sigma_{44}^k\, \rho$  and $(S_0^{j,\text{eff}} \sigma_{33}^j)\, \sigma_{44}^k\, \rho\, X_{lm} $ for the first perturbation in~\eqref{eq:NI2ndorder0} due to $V_{jk}^{34}$.  When  $l,m\neq j$ these terms decay to $0$, since $\P_0[(S_0^{j,\text{eff}} \sigma_{33}^j)\,X_{lm}\, \sigma_{44}^k\, \rho]= \lim_{t\rightarrow\infty}e^{-it (H_0^{j,\text{eff}}+H_0^{l,\text{eff}}+H_0^{m,\text{eff}})}(S_0^{j,\text{eff}} \sigma_{33}^j)\,X_{lm}\, \sigma_{44}^k\, \rho= (P_0^j S_0^{j,\text{eff}} \sigma_{33}^j) (P_0^l P_0^m \, X_{lm}\,\sigma_{44}^k )\, \rho =0$ as $P_0^j S_0^{j,\text{eff}}=0$. Similarly,  $\P_0[(S_0^{j,\text{eff}} \sigma_{33}^j)\, \sigma_{44}^k\, \rho\, X_{lm}]= (P_0^j S_0^{j,\text{eff}} \sigma_{33}^j) \sigma_{44}^k\, \rho (X_{lm} P_0^l P_0^m)= 0$. Analogously, such terms are 0 for the first perturbation in Eq.~\eqref{eq:NI2ndorder0} due to the exchange interaction or $\delta_3$-detuning.

\end{appendix}

\begin{thebibliography}{46}%
	\makeatletter
	\providecommand \@ifxundefined [1]{%
		\@ifx{#1\undefined}
	}%
	\providecommand \@ifnum [1]{%
		\ifnum #1\expandafter \@firstoftwo
		\else \expandafter \@secondoftwo
		\fi
	}%
	\providecommand \@ifx [1]{%
		\ifx #1\expandafter \@firstoftwo
		\else \expandafter \@secondoftwo
		\fi
	}%
	\providecommand \natexlab [1]{#1}%
	\providecommand \enquote  [1]{``#1''}%
	\providecommand \bibnamefont  [1]{#1}%
	\providecommand \bibfnamefont [1]{#1}%
	\providecommand \citenamefont [1]{#1}%
	\providecommand \href@noop [0]{\@secondoftwo}%
	\providecommand \href [0]{\begingroup \@sanitize@url \@href}%
	\providecommand \@href[1]{\@@startlink{#1}\@@href}%
	\providecommand \@@href[1]{\endgroup#1\@@endlink}%
	\providecommand \@sanitize@url [0]{\catcode `\\12\catcode `\$12\catcode
		`\&12\catcode `\#12\catcode `\^12\catcode `\_12\catcode `\%12\relax}%
	\providecommand \@@startlink[1]{}%
	\providecommand \@@endlink[0]{}%
	\providecommand \url  [0]{\begingroup\@sanitize@url \@url }%
	\providecommand \@url [1]{\endgroup\@href {#1}{\urlprefix }}%
	\providecommand \urlprefix  [0]{URL }%
	\providecommand \Eprint [0]{\href }%
	\providecommand \doibase [0]{http://dx.doi.org/}%
	\providecommand \selectlanguage [0]{\@gobble}%
	\providecommand \bibinfo  [0]{\@secondoftwo}%
	\providecommand \bibfield  [0]{\@secondoftwo}%
	\providecommand \translation [1]{[#1]}%
	\providecommand \BibitemOpen [0]{}%
	\providecommand \bibitemStop [0]{}%
	\providecommand \bibitemNoStop [0]{.\EOS\space}%
	\providecommand \EOS [0]{\spacefactor3000\relax}%
	\providecommand \BibitemShut  [1]{\csname bibitem#1\endcsname}%
	\let\auto@bib@innerbib\@empty
	\bibitem [{\citenamefont {Fleischhauer}\ \emph {et~al.}(2005)\citenamefont
		{Fleischhauer}, \citenamefont {Imamoglu},\ and\ \citenamefont
		{Marangos}}]{fleischhauer_electromagnetically_2005}%
	\BibitemOpen
	\bibfield  {author} {\bibinfo {author} {\bibfnamefont {M.}~\bibnamefont
			{Fleischhauer}}, \bibinfo {author} {\bibfnamefont {A.}~\bibnamefont
			{Imamoglu}}, \ and\ \bibinfo {author} {\bibfnamefont {J.~P.}\ \bibnamefont
			{Marangos}},\ }\href {\doibase 10.1103/RevModPhys.77.633} {\bibfield
		{journal} {\bibinfo  {journal} {Rev. Mod. Phys.}\ }\textbf {\bibinfo {volume}
			{77}},\ \bibinfo {pages} {633} (\bibinfo {year} {2005})}\BibitemShut
	{NoStop}%
	\bibitem [{\citenamefont {Firstenberg}\ \emph {et~al.}(2016)\citenamefont
		{Firstenberg}, \citenamefont {Adams},\ and\ \citenamefont
		{Hofferberth}}]{Hofferberth2016d}%
	\BibitemOpen
	\bibfield  {author} {\bibinfo {author} {\bibfnamefont {O.}~\bibnamefont
			{Firstenberg}}, \bibinfo {author} {\bibfnamefont {C.~S.}\ \bibnamefont
			{Adams}}, \ and\ \bibinfo {author} {\bibfnamefont {S.}~\bibnamefont
			{Hofferberth}},\ }\href {\doibase 10.1088/0953-4075/49/15/152003} {\bibfield
		{journal} {\bibinfo  {journal} {J. Phys. B}\ }\textbf {\bibinfo {volume}
			{49}},\ \bibinfo {pages} {152003} (\bibinfo {year} {2016})}\BibitemShut
	{NoStop}%
	\bibitem [{\citenamefont {Murray}\ and\ \citenamefont
		{Pohl}(2016)}]{murray_review_2016}%
	\BibitemOpen
	\bibfield  {author} {\bibinfo {author} {\bibfnamefont {C.}~\bibnamefont
			{Murray}}\ and\ \bibinfo {author} {\bibfnamefont {T.}~\bibnamefont {Pohl}},\
	}\href {\doibase https://doi.org/10.1016/bs.aamop.2016.04.005} {\bibfield
		{journal} {\bibinfo  {journal} {Adv. In Atomic, Molecular, and Optical
				Physics}\ }\textbf {\bibinfo {volume} {65}},\ \bibinfo {pages} {321 }
		(\bibinfo {year} {2016})}\BibitemShut {NoStop}%
	\bibitem [{\citenamefont {Li}\ and\ \citenamefont
		{Kuzmich}(2016)}]{li_quantum_2016}%
	\BibitemOpen
	\bibfield  {author} {\bibinfo {author} {\bibfnamefont {L.}~\bibnamefont
			{Li}}\ and\ \bibinfo {author} {\bibfnamefont {A.}~\bibnamefont {Kuzmich}},\
	}\href {\doibase 10.1038/ncomms13618} {\bibfield  {journal} {\bibinfo
			{journal} {Nat. Comm.}\ }\textbf {\bibinfo {volume} {7}},\ \bibinfo {pages}
		{13618} (\bibinfo {year} {2016})}\BibitemShut {NoStop}%
	\bibitem [{\citenamefont {Distante}\ \emph {et~al.}(2017)\citenamefont
		{Distante}, \citenamefont {Farrera}, \citenamefont {Padrón-Brito},
		\citenamefont {Paredes-Barato}, \citenamefont {Heinze},\ and\ \citenamefont
		{{de Riedmatten}}}]{distante_storing_2017}%
	\BibitemOpen
	\bibfield  {author} {\bibinfo {author} {\bibfnamefont {E.}~\bibnamefont
			{Distante}}, \bibinfo {author} {\bibfnamefont {P.}~\bibnamefont {Farrera}},
		\bibinfo {author} {\bibfnamefont {A.}~\bibnamefont {Padrón-Brito}}, \bibinfo
		{author} {\bibfnamefont {D.}~\bibnamefont {Paredes-Barato}}, \bibinfo
		{author} {\bibfnamefont {G.}~\bibnamefont {Heinze}}, \ and\ \bibinfo {author}
		{\bibfnamefont {H.}~\bibnamefont {{de Riedmatten}}},\ }\href {\doibase
		10.1038/ncomms14072} {\bibfield  {journal} {\bibinfo  {journal} {Nat. Comm.}\
		}\textbf {\bibinfo {volume} {8}},\ \bibinfo {pages} {14072} (\bibinfo {year}
		{2017})}\BibitemShut {NoStop}%
	\bibitem [{\citenamefont {Boller}\ \emph {et~al.}(1991)\citenamefont {Boller},
		\citenamefont {Imamoglu},\ and\ \citenamefont
		{Harris}}]{boller_observation_1991}%
	\BibitemOpen
	\bibfield  {author} {\bibinfo {author} {\bibfnamefont {K.-J.}\ \bibnamefont
			{Boller}}, \bibinfo {author} {\bibfnamefont {A.}~\bibnamefont {Imamoglu}}, \
		and\ \bibinfo {author} {\bibfnamefont {S.~E.}\ \bibnamefont {Harris}},\
	}\href {\doibase 10.1103/PhysRevLett.66.2593} {\bibfield  {journal} {\bibinfo
			{journal} {Phys. Rev. Lett.}\ }\textbf {\bibinfo {volume} {66}},\ \bibinfo
		{pages} {2593} (\bibinfo {year} {1991})}\BibitemShut {NoStop}%
	\bibitem [{\citenamefont {Pritchard}\ \emph {et~al.}(2010)\citenamefont
		{Pritchard}, \citenamefont {Maxwell}, \citenamefont {Gauguet}, \citenamefont
		{Weatherill}, \citenamefont {Jones},\ and\ \citenamefont
		{Adams}}]{Adams2010}%
	\BibitemOpen
	\bibfield  {author} {\bibinfo {author} {\bibfnamefont {J.~D.}\ \bibnamefont
			{Pritchard}}, \bibinfo {author} {\bibfnamefont {D.}~\bibnamefont {Maxwell}},
		\bibinfo {author} {\bibfnamefont {A.}~\bibnamefont {Gauguet}}, \bibinfo
		{author} {\bibfnamefont {K.~J.}\ \bibnamefont {Weatherill}}, \bibinfo
		{author} {\bibfnamefont {M.~P.~A.}\ \bibnamefont {Jones}}, \ and\ \bibinfo
		{author} {\bibfnamefont {C.~S.}\ \bibnamefont {Adams}},\ }\href {\doibase
		10.1103/PhysRevLett.105.193603} {\bibfield  {journal} {\bibinfo  {journal}
			{Phys. Rev. Lett.}\ }\textbf {\bibinfo {volume} {105}},\ \bibinfo {pages}
		{193603} (\bibinfo {year} {2010})}\BibitemShut {NoStop}%
	\bibitem [{\citenamefont {Maxwell}\ \emph {et~al.}(2013)\citenamefont
		{Maxwell}, \citenamefont {Szwer}, \citenamefont {Paredes-Barato},
		\citenamefont {Busche}, \citenamefont {Pritchard}, \citenamefont {Gauguet},
		\citenamefont {Weatherill}, \citenamefont {Jones},\ and\ \citenamefont
		{Adams}}]{Adams2013}%
	\BibitemOpen
	\bibfield  {author} {\bibinfo {author} {\bibfnamefont {D.}~\bibnamefont
			{Maxwell}}, \bibinfo {author} {\bibfnamefont {D.~J.}\ \bibnamefont {Szwer}},
		\bibinfo {author} {\bibfnamefont {D.}~\bibnamefont {Paredes-Barato}},
		\bibinfo {author} {\bibfnamefont {H.}~\bibnamefont {Busche}}, \bibinfo
		{author} {\bibfnamefont {J.~D.}\ \bibnamefont {Pritchard}}, \bibinfo {author}
		{\bibfnamefont {A.}~\bibnamefont {Gauguet}}, \bibinfo {author} {\bibfnamefont
			{K.~J.}\ \bibnamefont {Weatherill}}, \bibinfo {author} {\bibfnamefont
			{M.~P.~A.}\ \bibnamefont {Jones}}, \ and\ \bibinfo {author} {\bibfnamefont
			{C.~S.}\ \bibnamefont {Adams}},\ }\href
	{http://link.aps.org/doi/10.1103/PhysRevLett.110.103001} {\bibfield
		{journal} {\bibinfo  {journal} {Phys. Rev. Lett.}\ }\textbf {\bibinfo
			{volume} {110}},\ \bibinfo {pages} {103001} (\bibinfo {year}
		{2013})}\BibitemShut {NoStop}%
	\bibitem [{\citenamefont {Parigi}\ \emph {et~al.}(2012)\citenamefont {Parigi},
		\citenamefont {Bimbard}, \citenamefont {Stanojevic}, \citenamefont
		{Hilliard}, \citenamefont {Nogrette}, \citenamefont {Tualle-Brouri},
		\citenamefont {Ourjoumtsev},\ and\ \citenamefont {Grangier}}]{Grangier2012}%
	\BibitemOpen
	\bibfield  {author} {\bibinfo {author} {\bibfnamefont {V.}~\bibnamefont
			{Parigi}}, \bibinfo {author} {\bibfnamefont {E.}~\bibnamefont {Bimbard}},
		\bibinfo {author} {\bibfnamefont {J.}~\bibnamefont {Stanojevic}}, \bibinfo
		{author} {\bibfnamefont {A.~J.}\ \bibnamefont {Hilliard}}, \bibinfo {author}
		{\bibfnamefont {F.}~\bibnamefont {Nogrette}}, \bibinfo {author}
		{\bibfnamefont {R.}~\bibnamefont {Tualle-Brouri}}, \bibinfo {author}
		{\bibfnamefont {A.}~\bibnamefont {Ourjoumtsev}}, \ and\ \bibinfo {author}
		{\bibfnamefont {P.}~\bibnamefont {Grangier}},\ }\href {\doibase
		10.1103/PhysRevLett.109.233602} {\bibfield  {journal} {\bibinfo  {journal}
			{Phys. Rev. Lett.}\ }\textbf {\bibinfo {volume} {109}},\ \bibinfo {pages}
		{233602} (\bibinfo {year} {2012})}\BibitemShut {NoStop}%
	\bibitem [{\citenamefont {Dudin}\ and\ \citenamefont
		{Kuzmich}(2012)}]{Kuzmich2012b}%
	\BibitemOpen
	\bibfield  {author} {\bibinfo {author} {\bibfnamefont {Y.~O.}\ \bibnamefont
			{Dudin}}\ and\ \bibinfo {author} {\bibfnamefont {A.}~\bibnamefont
			{Kuzmich}},\ }\href {\doibase 10.1126/science.1217901} {\bibfield  {journal}
		{\bibinfo  {journal} {Science}\ }\textbf {\bibinfo {volume} {336}},\ \bibinfo
		{pages} {887} (\bibinfo {year} {2012})}\BibitemShut {NoStop}%
	\bibitem [{\citenamefont {Li}\ \emph {et~al.}(2013)\citenamefont {Li},
		\citenamefont {Dudin},\ and\ \citenamefont {Kuzmich}}]{Kuzmich2013}%
	\BibitemOpen
	\bibfield  {author} {\bibinfo {author} {\bibfnamefont {L.}~\bibnamefont
			{Li}}, \bibinfo {author} {\bibfnamefont {Y.~O.}\ \bibnamefont {Dudin}}, \
		and\ \bibinfo {author} {\bibfnamefont {A.}~\bibnamefont {Kuzmich}},\ }\href
	{\doibase 10.1038/nature12227} {\bibfield  {journal} {\bibinfo  {journal}
			{Nature}\ }\textbf {\bibinfo {volume} {498}},\ \bibinfo {pages} {466}
		(\bibinfo {year} {2013})}\BibitemShut {NoStop}%
	\bibitem [{\citenamefont {Peyronel}\ \emph {et~al.}(2012)\citenamefont
		{Peyronel}, \citenamefont {Firstenberg}, \citenamefont {Liang}, \citenamefont
		{Hofferberth}, \citenamefont {Gorshkov}, \citenamefont {Pohl}, \citenamefont
		{Lukin},\ and\ \citenamefont {Vuleti\'c}}]{Vuletic2012}%
	\BibitemOpen
	\bibfield  {author} {\bibinfo {author} {\bibfnamefont {T.}~\bibnamefont
			{Peyronel}}, \bibinfo {author} {\bibfnamefont {O.}~\bibnamefont
			{Firstenberg}}, \bibinfo {author} {\bibfnamefont {Q.}~\bibnamefont {Liang}},
		\bibinfo {author} {\bibfnamefont {S.}~\bibnamefont {Hofferberth}}, \bibinfo
		{author} {\bibfnamefont {A.}~\bibnamefont {Gorshkov}}, \bibinfo {author}
		{\bibfnamefont {T.}~\bibnamefont {Pohl}}, \bibinfo {author} {\bibfnamefont
			{M.}~\bibnamefont {Lukin}}, \ and\ \bibinfo {author} {\bibfnamefont
			{V.}~\bibnamefont {Vuleti\'c}},\ }\href {\doibase 10.1038/nature11361}
	{\bibfield  {journal} {\bibinfo  {journal} {Nature}\ }\textbf {\bibinfo
			{volume} {488}},\ \bibinfo {pages} {57} (\bibinfo {year} {2012})}\BibitemShut
	{NoStop}%
	\bibitem [{\citenamefont {Firstenberg}\ \emph {et~al.}(2013)\citenamefont
		{Firstenberg}, \citenamefont {Peyronel}, \citenamefont {Liang}, \citenamefont
		{Gorshkov}, \citenamefont {Lukin},\ and\ \citenamefont
		{Vuleti\'c}}]{Vuletic2013b}%
	\BibitemOpen
	\bibfield  {author} {\bibinfo {author} {\bibfnamefont {O.}~\bibnamefont
			{Firstenberg}}, \bibinfo {author} {\bibfnamefont {T.}~\bibnamefont
			{Peyronel}}, \bibinfo {author} {\bibfnamefont {Q.}~\bibnamefont {Liang}},
		\bibinfo {author} {\bibfnamefont {A.~V.}\ \bibnamefont {Gorshkov}}, \bibinfo
		{author} {\bibfnamefont {M.~D.}\ \bibnamefont {Lukin}}, \ and\ \bibinfo
		{author} {\bibfnamefont {V.}~\bibnamefont {Vuleti\'c}},\ }\href {\doibase
		doi:10.1038/nature12512} {\bibfield  {journal} {\bibinfo  {journal} {Nature}\
		}\textbf {\bibinfo {volume} {502}},\ \bibinfo {pages} {71} (\bibinfo {year}
		{2013})}\BibitemShut {NoStop}%
	\bibitem [{\citenamefont {Baur}\ \emph {et~al.}(2014)\citenamefont {Baur},
		\citenamefont {Tiarks}, \citenamefont {Rempe},\ and\ \citenamefont
		{D\"urr}}]{Baur_photon_switch}%
	\BibitemOpen
	\bibfield  {author} {\bibinfo {author} {\bibfnamefont {S.}~\bibnamefont
			{Baur}}, \bibinfo {author} {\bibfnamefont {D.}~\bibnamefont {Tiarks}},
		\bibinfo {author} {\bibfnamefont {G.}~\bibnamefont {Rempe}}, \ and\ \bibinfo
		{author} {\bibfnamefont {S.}~\bibnamefont {D\"urr}},\ }\href {\doibase
		10.1103/PhysRevLett.112.073901} {\bibfield  {journal} {\bibinfo  {journal}
			{Phys. Rev. Lett.}\ }\textbf {\bibinfo {volume} {112}},\ \bibinfo {pages}
		{073901} (\bibinfo {year} {2014})}\BibitemShut {NoStop}%
	\bibitem [{\citenamefont {Gorniaczyk}\ \emph {et~al.}(2014)\citenamefont
		{Gorniaczyk}, \citenamefont {Tresp}, \citenamefont {Schmidt}, \citenamefont
		{Fedder},\ and\ \citenamefont {Hofferberth}}]{gorniaczyk_single-photon_2014}%
	\BibitemOpen
	\bibfield  {author} {\bibinfo {author} {\bibfnamefont {H.}~\bibnamefont
			{Gorniaczyk}}, \bibinfo {author} {\bibfnamefont {C.}~\bibnamefont {Tresp}},
		\bibinfo {author} {\bibfnamefont {J.}~\bibnamefont {Schmidt}}, \bibinfo
		{author} {\bibfnamefont {H.}~\bibnamefont {Fedder}}, \ and\ \bibinfo {author}
		{\bibfnamefont {S.}~\bibnamefont {Hofferberth}},\ }\href {\doibase
		10.1103/PhysRevLett.113.053601} {\bibfield  {journal} {\bibinfo  {journal}
			{Phys. Rev. Lett.}\ }\textbf {\bibinfo {volume} {113}},\ \bibinfo {pages}
		{053601} (\bibinfo {year} {2014})}\BibitemShut {NoStop}%
	\bibitem [{\citenamefont {Tiarks}\ \emph {et~al.}(2014)\citenamefont {Tiarks},
		\citenamefont {Baur}, \citenamefont {Schneider}, \citenamefont {D\"urr},\
		and\ \citenamefont {Rempe}}]{tiarks_photon_transistor_2014}%
	\BibitemOpen
	\bibfield  {author} {\bibinfo {author} {\bibfnamefont {D.}~\bibnamefont
			{Tiarks}}, \bibinfo {author} {\bibfnamefont {S.}~\bibnamefont {Baur}},
		\bibinfo {author} {\bibfnamefont {K.}~\bibnamefont {Schneider}}, \bibinfo
		{author} {\bibfnamefont {S.}~\bibnamefont {D\"urr}}, \ and\ \bibinfo {author}
		{\bibfnamefont {G.}~\bibnamefont {Rempe}},\ }\href {\doibase
		10.1103/PhysRevLett.113.053602} {\bibfield  {journal} {\bibinfo  {journal}
			{Phys. Rev. Lett.}\ }\textbf {\bibinfo {volume} {113}},\ \bibinfo {pages}
		{053602} (\bibinfo {year} {2014})}\BibitemShut {NoStop}%
	\bibitem [{\citenamefont {Gorniaczyk}\ \emph {et~al.}(2016)\citenamefont
		{Gorniaczyk}, \citenamefont {Tresp}, \citenamefont {Bienias}, \citenamefont
		{Paris-Mandoki}, \citenamefont {Li}, \citenamefont {Mirgorodskiy},
		\citenamefont {B{\"u}chler}, \citenamefont {Lesanovsky},\ and\ \citenamefont
		{Hofferberth}}]{gorniaczyk_enhancement_2016}%
	\BibitemOpen
	\bibfield  {author} {\bibinfo {author} {\bibfnamefont {H.}~\bibnamefont
			{Gorniaczyk}}, \bibinfo {author} {\bibfnamefont {C.}~\bibnamefont {Tresp}},
		\bibinfo {author} {\bibfnamefont {P.}~\bibnamefont {Bienias}}, \bibinfo
		{author} {\bibfnamefont {A.}~\bibnamefont {Paris-Mandoki}}, \bibinfo {author}
		{\bibfnamefont {W.}~\bibnamefont {Li}}, \bibinfo {author} {\bibfnamefont
			{I.}~\bibnamefont {Mirgorodskiy}}, \bibinfo {author} {\bibfnamefont {H.~P.}\
			\bibnamefont {B{\"u}chler}}, \bibinfo {author} {\bibfnamefont
			{I.}~\bibnamefont {Lesanovsky}}, \ and\ \bibinfo {author} {\bibfnamefont
			{S.}~\bibnamefont {Hofferberth}},\ }\href {\doibase 10.1038/ncomms12480}
	{\bibfield  {journal} {\bibinfo  {journal} {Nat. Comm.}\ }\textbf {\bibinfo
			{volume} {7}},\ \bibinfo {pages} {12480} (\bibinfo {year}
		{2016})}\BibitemShut {NoStop}%
	\bibitem [{\citenamefont {Li}\ and\ \citenamefont
		{Lesanovsky}(2015{\natexlab{a}})}]{li_coherence_2015}%
	\BibitemOpen
	\bibfield  {author} {\bibinfo {author} {\bibfnamefont {W.}~\bibnamefont
			{Li}}\ and\ \bibinfo {author} {\bibfnamefont {I.}~\bibnamefont
			{Lesanovsky}},\ }\href {\doibase 10.1103/PhysRevA.92.043828} {\bibfield
		{journal} {\bibinfo  {journal} {Phys. Rev. A}\ }\textbf {\bibinfo {volume}
			{92}},\ \bibinfo {pages} {043828} (\bibinfo {year}
		{2015}{\natexlab{a}})}\BibitemShut {NoStop}%
	\bibitem [{\citenamefont {Murray}\ \emph {et~al.}(2016)\citenamefont {Murray},
		\citenamefont {Gorshkov},\ and\ \citenamefont
		{Pohl}}]{murray_many-body_2016}%
	\BibitemOpen
	\bibfield  {author} {\bibinfo {author} {\bibfnamefont {C.~R.}\ \bibnamefont
			{Murray}}, \bibinfo {author} {\bibfnamefont {A.~V.}\ \bibnamefont
			{Gorshkov}}, \ and\ \bibinfo {author} {\bibfnamefont {T.}~\bibnamefont
			{Pohl}},\ }\href {\doibase 10.1088/1367-2630/18/9/092001} {\bibfield
		{journal} {\bibinfo  {journal} {New J. Phys.}\ }\textbf {\bibinfo {volume}
			{18}},\ \bibinfo {pages} {092001} (\bibinfo {year} {2016})}\BibitemShut
	{NoStop}%
	\bibitem [{\citenamefont {Tresp}\ \emph {et~al.}(2016)\citenamefont {Tresp},
		\citenamefont {Zimmer}, \citenamefont {Mirgorodskiy}, \citenamefont
		{Gorniaczyk}, \citenamefont {Paris-Mandoki},\ and\ \citenamefont
		{Hofferberth}}]{tresp_single-photon_2016}%
	\BibitemOpen
	\bibfield  {author} {\bibinfo {author} {\bibfnamefont {C.}~\bibnamefont
			{Tresp}}, \bibinfo {author} {\bibfnamefont {C.}~\bibnamefont {Zimmer}},
		\bibinfo {author} {\bibfnamefont {I.}~\bibnamefont {Mirgorodskiy}}, \bibinfo
		{author} {\bibfnamefont {H.}~\bibnamefont {Gorniaczyk}}, \bibinfo {author}
		{\bibfnamefont {A.}~\bibnamefont {Paris-Mandoki}}, \ and\ \bibinfo {author}
		{\bibfnamefont {S.}~\bibnamefont {Hofferberth}},\ }\href {\doibase
		10.1103/PhysRevLett.117.223001} {\bibfield  {journal} {\bibinfo  {journal}
			{Phys. Rev. Lett.}\ }\textbf {\bibinfo {volume} {117}},\ \bibinfo {pages}
		{223001} (\bibinfo {year} {2016})}\BibitemShut {NoStop}%
	\bibitem [{\citenamefont {Gorshkov}\ \emph {et~al.}(2011)\citenamefont
		{Gorshkov}, \citenamefont {Otterbach}, \citenamefont {Fleischhauer},
		\citenamefont {Pohl},\ and\ \citenamefont
		{Lukin}}]{gorshkov_photon-photon_2011}%
	\BibitemOpen
	\bibfield  {author} {\bibinfo {author} {\bibfnamefont {A.~V.}\ \bibnamefont
			{Gorshkov}}, \bibinfo {author} {\bibfnamefont {J.}~\bibnamefont {Otterbach}},
		\bibinfo {author} {\bibfnamefont {M.}~\bibnamefont {Fleischhauer}}, \bibinfo
		{author} {\bibfnamefont {T.}~\bibnamefont {Pohl}}, \ and\ \bibinfo {author}
		{\bibfnamefont {M.~D.}\ \bibnamefont {Lukin}},\ }\href {\doibase
		10.1103/PhysRevLett.107.133602} {\bibfield  {journal} {\bibinfo  {journal}
			{Phys. Rev. Lett.}\ }\textbf {\bibinfo {volume} {107}},\ \bibinfo {pages}
		{133602} (\bibinfo {year} {2011})}\BibitemShut {NoStop}%
	\bibitem [{\citenamefont {Macieszczak}\ \emph {et~al.}(2016)\citenamefont
		{Macieszczak}, \citenamefont {Gu\ifmmode \mbox{\c{t}}\else
			\c{t}\fi{}\ifmmode~\u{a}\else \u{a}\fi{}}, \citenamefont {Lesanovsky},\ and\
		\citenamefont {Garrahan}}]{macieszczak_metastability_2016}%
	\BibitemOpen
	\bibfield  {author} {\bibinfo {author} {\bibfnamefont {K.}~\bibnamefont
			{Macieszczak}}, \bibinfo {author} {\bibfnamefont {M.}~\bibnamefont
			{Gu\ifmmode \mbox{\c{t}}\else \c{t}\fi{}\ifmmode~\u{a}\else \u{a}\fi{}}},
		\bibinfo {author} {\bibfnamefont {I.}~\bibnamefont {Lesanovsky}}, \ and\
		\bibinfo {author} {\bibfnamefont {J.~P.}\ \bibnamefont {Garrahan}},\ }\href
	{\doibase 10.1103/PhysRevLett.116.240404} {\bibfield  {journal} {\bibinfo
			{journal} {Phys. Rev. Lett.}\ }\textbf {\bibinfo {volume} {116}},\ \bibinfo
		{pages} {240404} (\bibinfo {year} {2016})}\BibitemShut {NoStop}%
	\bibitem [{\citenamefont {Gaul}\ \emph {et~al.}(2016)\citenamefont {Gaul},
		\citenamefont {DeSalvo}, \citenamefont {Aman}, \citenamefont {Dunning},
		\citenamefont {Killian},\ and\ \citenamefont
		{Pohl}}]{coherent_dissipative_interaction}%
	\BibitemOpen
	\bibfield  {author} {\bibinfo {author} {\bibfnamefont {C.}~\bibnamefont
			{Gaul}}, \bibinfo {author} {\bibfnamefont {B.~J.}\ \bibnamefont {DeSalvo}},
		\bibinfo {author} {\bibfnamefont {J.~A.}\ \bibnamefont {Aman}}, \bibinfo
		{author} {\bibfnamefont {F.~B.}\ \bibnamefont {Dunning}}, \bibinfo {author}
		{\bibfnamefont {T.~C.}\ \bibnamefont {Killian}}, \ and\ \bibinfo {author}
		{\bibfnamefont {T.}~\bibnamefont {Pohl}},\ }\href {\doibase
		10.1103/PhysRevLett.116.243001} {\bibfield  {journal} {\bibinfo  {journal}
			{Phys. Rev. Lett.}\ }\textbf {\bibinfo {volume} {116}},\ \bibinfo {pages}
		{243001} (\bibinfo {year} {2016})}\BibitemShut {NoStop}%
	\bibitem [{\citenamefont {Arimondo}(1996)}]{Arimondo1996}%
	\BibitemOpen
	\bibfield  {author} {\bibinfo {author} {\bibfnamefont {E.}~\bibnamefont
			{Arimondo}},\ }\href {\doibase
		http://dx.doi.org/10.1016/S0079-6638(08)70531-6} {\bibfield  {journal}
		{\bibinfo  {journal} {Progress in Optics}\ }\textbf {\bibinfo {volume}
			{35}},\ \bibinfo {pages} {257 } (\bibinfo {year} {1996})}\BibitemShut
	{NoStop}%
	\bibitem [{\citenamefont {Harris}(1997)}]{Harris1997}%
	\BibitemOpen
	\bibfield  {author} {\bibinfo {author} {\bibfnamefont {S.~E.}\ \bibnamefont
			{Harris}},\ }\href {\doibase http://dx.doi.org/10.1063/1.881806} {\bibfield
		{journal} {\bibinfo  {journal} {Physics Today}\ }\textbf {\bibinfo {volume}
			{50}},\ \bibinfo {pages} {36} (\bibinfo {year} {1997})}\BibitemShut {NoStop}%
	\bibitem [{\citenamefont {Diehl}\ \emph {et~al.}(2008)\citenamefont {Diehl},
		\citenamefont {Micheli}, \citenamefont {Kantian}, \citenamefont {Kraus},
		\citenamefont {B{\"{u}}chler},\ and\ \citenamefont
		{Zoller}}]{diehl_quantum_2008}%
	\BibitemOpen
	\bibfield  {author} {\bibinfo {author} {\bibfnamefont {S.}~\bibnamefont
			{Diehl}}, \bibinfo {author} {\bibfnamefont {A.}~\bibnamefont {Micheli}},
		\bibinfo {author} {\bibfnamefont {A.}~\bibnamefont {Kantian}}, \bibinfo
		{author} {\bibfnamefont {B.}~\bibnamefont {Kraus}}, \bibinfo {author}
		{\bibfnamefont {H.~P.}\ \bibnamefont {B{\"{u}}chler}}, \ and\ \bibinfo
		{author} {\bibfnamefont {P.}~\bibnamefont {Zoller}},\ }\href {\doibase
		10.1038/nphys1073} {\bibfield  {journal} {\bibinfo  {journal} {Nat. Phys.}\
		}\textbf {\bibinfo {volume} {4}},\ \bibinfo {pages} {878} (\bibinfo {year}
		{2008})}\BibitemShut {NoStop}%
	\bibitem [{\citenamefont {Kraus}\ \emph {et~al.}(2008)\citenamefont {Kraus},
		\citenamefont {B\"uchler}, \citenamefont {Diehl}, \citenamefont {Kantian},
		\citenamefont {Micheli},\ and\ \citenamefont
		{Zoller}}]{kraus_preparation_2008}%
	\BibitemOpen
	\bibfield  {author} {\bibinfo {author} {\bibfnamefont {B.}~\bibnamefont
			{Kraus}}, \bibinfo {author} {\bibfnamefont {H.~P.}\ \bibnamefont
			{B\"uchler}}, \bibinfo {author} {\bibfnamefont {S.}~\bibnamefont {Diehl}},
		\bibinfo {author} {\bibfnamefont {A.}~\bibnamefont {Kantian}}, \bibinfo
		{author} {\bibfnamefont {A.}~\bibnamefont {Micheli}}, \ and\ \bibinfo
		{author} {\bibfnamefont {P.}~\bibnamefont {Zoller}},\ }\href {\doibase
		10.1103/PhysRevA.78.042307} {\bibfield  {journal} {\bibinfo  {journal} {Phys.
				Rev. A}\ }\textbf {\bibinfo {volume} {78}},\ \bibinfo {pages} {042307}
		(\bibinfo {year} {2008})}\BibitemShut {NoStop}%
	\bibitem [{\citenamefont {Zanardi}\ and\ \citenamefont
		{Campos~Venuti}(2014)}]{zanardi_coherent_2014}%
	\BibitemOpen
	\bibfield  {author} {\bibinfo {author} {\bibfnamefont {P.}~\bibnamefont
			{Zanardi}}\ and\ \bibinfo {author} {\bibfnamefont {L.}~\bibnamefont
			{Campos~Venuti}},\ }\href {\doibase 10.1103/PhysRevLett.113.240406}
	{\bibfield  {journal} {\bibinfo  {journal} {Phys. Rev. Lett.}\ }\textbf
		{\bibinfo {volume} {113}},\ \bibinfo {pages} {240406} (\bibinfo {year}
		{2014})}\BibitemShut {NoStop}%
	\bibitem [{\citenamefont {Zanardi}\ and\ \citenamefont
		{Campos~Venuti}(2015)}]{zanardi_geometry_2015}%
	\BibitemOpen
	\bibfield  {author} {\bibinfo {author} {\bibfnamefont {P.}~\bibnamefont
			{Zanardi}}\ and\ \bibinfo {author} {\bibfnamefont {L.}~\bibnamefont
			{Campos~Venuti}},\ }\href {\doibase 10.1103/PhysRevA.91.052324} {\bibfield
		{journal} {\bibinfo  {journal} {Phys. Rev. A}\ }\textbf {\bibinfo {volume}
			{91}},\ \bibinfo {pages} {052324} (\bibinfo {year} {2015})}\BibitemShut
	{NoStop}%
	\bibitem [{\citenamefont {Tiarks}\ \emph {et~al.}(2016)\citenamefont {Tiarks},
		\citenamefont {Schmidt}, \citenamefont {Rempe},\ and\ \citenamefont
		{D\"urr}}]{tiarks_optical_2016}%
	\BibitemOpen
	\bibfield  {author} {\bibinfo {author} {\bibfnamefont {D.}~\bibnamefont
			{Tiarks}}, \bibinfo {author} {\bibfnamefont {S.}~\bibnamefont {Schmidt}},
		\bibinfo {author} {\bibfnamefont {G.}~\bibnamefont {Rempe}}, \ and\ \bibinfo
		{author} {\bibfnamefont {S.}~\bibnamefont {D\"urr}},\ }\href {\doibase
		http://dx.doi.org/10.1126/sciadv.1600036} {\bibfield  {journal} {\bibinfo
			{journal} {Sci. Adv.}\ }\textbf {\bibinfo {volume} {2}},\ \bibinfo {pages}
		{e1600036} (\bibinfo {year} {2016})}\BibitemShut {NoStop}%
	\bibitem [{\citenamefont {Thompson}\ \emph {et~al.}(2017)\citenamefont
		{Thompson}, \citenamefont {Nicholson}, \citenamefont {Liang}, \citenamefont
		{Cantu}, \citenamefont {Venkatramani}, \citenamefont {Choi}, \citenamefont
		{Fedorov}, \citenamefont {Viscor}, \citenamefont {Pohl}, \citenamefont
		{Lukin},\ and\ \citenamefont {Vuleti{\'c}}}]{thomson_symmetry_2017}%
	\BibitemOpen
	\bibfield  {author} {\bibinfo {author} {\bibfnamefont {J.~D.}\ \bibnamefont
			{Thompson}}, \bibinfo {author} {\bibfnamefont {T.~L.}\ \bibnamefont
			{Nicholson}}, \bibinfo {author} {\bibfnamefont {Q.-Y.}\ \bibnamefont
			{Liang}}, \bibinfo {author} {\bibfnamefont {S.~H.}\ \bibnamefont {Cantu}},
		\bibinfo {author} {\bibfnamefont {A.~V.}\ \bibnamefont {Venkatramani}},
		\bibinfo {author} {\bibfnamefont {S.}~\bibnamefont {Choi}}, \bibinfo {author}
		{\bibfnamefont {I.~A.}\ \bibnamefont {Fedorov}}, \bibinfo {author}
		{\bibfnamefont {D.}~\bibnamefont {Viscor}}, \bibinfo {author} {\bibfnamefont
			{T.}~\bibnamefont {Pohl}}, \bibinfo {author} {\bibfnamefont {M.~D.}\
			\bibnamefont {Lukin}}, \ and\ \bibinfo {author} {\bibfnamefont
			{V.}~\bibnamefont {Vuleti{\'c}}},\ }\href
	{http://dx.doi.org/10.1038/nature20823} {\bibfield  {journal} {\bibinfo
			{journal} {Nature}\ }\textbf {\bibinfo {volume} {542}},\ \bibinfo {pages}
		{206} (\bibinfo {year} {2017})}\BibitemShut {NoStop}%
	\bibitem [{\citenamefont {Li}\ \emph {et~al.}(2014)\citenamefont {Li},
		\citenamefont {Viscor}, \citenamefont {Hofferberth},\ and\ \citenamefont
		{Lesanovsky}}]{li_electromagnetically_2014}%
	\BibitemOpen
	\bibfield  {author} {\bibinfo {author} {\bibfnamefont {W.}~\bibnamefont
			{Li}}, \bibinfo {author} {\bibfnamefont {D.}~\bibnamefont {Viscor}}, \bibinfo
		{author} {\bibfnamefont {S.}~\bibnamefont {Hofferberth}}, \ and\ \bibinfo
		{author} {\bibfnamefont {I.}~\bibnamefont {Lesanovsky}},\ }\href {\doibase
		10.1103/PhysRevLett.112.243601} {\bibfield  {journal} {\bibinfo  {journal}
			{Phys. Rev. Lett.}\ }\textbf {\bibinfo {volume} {112}},\ \bibinfo {pages}
		{243601} (\bibinfo {year} {2014})}\BibitemShut {NoStop}%
	\bibitem [{\citenamefont {Li}\ and\ \citenamefont
		{Lesanovsky}(2015{\natexlab{b}})}]{coherence_Li_2015}%
	\BibitemOpen
	\bibfield  {author} {\bibinfo {author} {\bibfnamefont {W.}~\bibnamefont
			{Li}}\ and\ \bibinfo {author} {\bibfnamefont {I.}~\bibnamefont
			{Lesanovsky}},\ }\href {\doibase 10.1103/PhysRevA.92.043828} {\bibfield
		{journal} {\bibinfo  {journal} {Phys. Rev. A}\ }\textbf {\bibinfo {volume}
			{92}},\ \bibinfo {pages} {043828} (\bibinfo {year}
		{2015}{\natexlab{b}})}\BibitemShut {NoStop}%
	\bibitem [{\citenamefont {Lindblad}(1976)}]{Lindblad1976}%
	\BibitemOpen
	\bibfield  {author} {\bibinfo {author} {\bibfnamefont {G.}~\bibnamefont
			{Lindblad}},\ }\href@noop {} {\bibfield  {journal} {\bibinfo  {journal}
			{Comm. Math. Phys}\ }\textbf {\bibinfo {volume} {48}},\ \bibinfo {pages}
		{119} (\bibinfo {year} {1976})}\BibitemShut {NoStop}%
	\bibitem [{\citenamefont {Gorini}\ \emph {et~al.}(1976)\citenamefont {Gorini},
		\citenamefont {Kossakowski},\ and\ \citenamefont {Sudarshan}}]{Gorini1976}%
	\BibitemOpen
	\bibfield  {author} {\bibinfo {author} {\bibfnamefont {V.}~\bibnamefont
			{Gorini}}, \bibinfo {author} {\bibfnamefont {A.}~\bibnamefont {Kossakowski}},
		\ and\ \bibinfo {author} {\bibfnamefont {E.~C.~G.}\ \bibnamefont
			{Sudarshan}},\ }\href@noop {} {\bibfield  {journal} {\bibinfo  {journal} {J.
				Mat. Phys.}\ }\textbf {\bibinfo {volume} {17}},\ \bibinfo {pages} {821}
		(\bibinfo {year} {1976})}\BibitemShut {NoStop}%
	\bibitem [{Note1()}]{Note1}%
	\BibitemOpen
	\bibinfo {note} {As $\delta _3\not =0$ perturbs a DFS, the long-time dynamics
		timescale $\tau $ is actually determined with non-dissipative relaxation time
		$\tau _0'\leq \tau _0$ given by the inverse of the imaginary gap of the
		effective Hamiltonian of the single-atom dynamics $\protect \mathcal {L}$ at
		$\delta _3=0$, i.e. $H_j-\protect \frac {i}{2}\sigma _{22}^{j}$ instead of
		$\tau _0$~\cite {albert_adiabatic_2016}, see Appendix~\ref
		{app:NI}.}\BibitemShut {Stop}%
	\bibitem [{\citenamefont {Zanardi}\ \emph {et~al.}(2016)\citenamefont
		{Zanardi}, \citenamefont {Marshall},\ and\ \citenamefont
		{Campos~Venuti}}]{zanardi_dissipative_2016}%
	\BibitemOpen
	\bibfield  {author} {\bibinfo {author} {\bibfnamefont {P.}~\bibnamefont
			{Zanardi}}, \bibinfo {author} {\bibfnamefont {J.}~\bibnamefont {Marshall}}, \
		and\ \bibinfo {author} {\bibfnamefont {L.}~\bibnamefont {Campos~Venuti}},\
	}\href {\doibase 10.1103/PhysRevA.93.022312} {\bibfield  {journal} {\bibinfo
			{journal} {Phys. Rev. A}\ }\textbf {\bibinfo {volume} {93}},\ \bibinfo
		{pages} {022312} (\bibinfo {year} {2016})}\BibitemShut {NoStop}%
	\bibitem [{Note2()}]{Note2}%
	\BibitemOpen
	\bibinfo {note} {As the probe field perturbs a DFS, the long-time dynamics
		timescales are actually determined by non-dissipative relaxation time $\tau
		_{0,\protect \mathrm {int}}'\leq \tau _{0,\protect \mathrm {int}}$, given by
		the inverse of the imaginary gap of the effective Hamiltonian of the dynamics
		$\protect \mathcal {L}$ at $\Omega _p=0$, i.e. $H-\protect \frac {i}{2}\DOTSB
		\sum@ \slimits@ _{j=1}^N\sigma _{22}^{j}$ instead of $\tau _{0,\protect
			\mathrm {int}}$, cf.~\cite {albert_adiabatic_2016}, see Appendix~\ref
		{app:Iderivation}.}\BibitemShut {Stop}%
	\bibitem [{Note3()}]{Note3}%
	\BibitemOpen
	\bibinfo {note} {This structure is not changed by higher order corrections,
		as all the eigenmodes of the full dynamics in Eq.~\protect \textup {\hbox
			{\mathsurround \z@ \protect \normalfont (\ignorespaces \ref
				{eq:master}\unskip \@@italiccorr )}} are separable, thus guaranteeing
		locality of the dynamics}\BibitemShut {NoStop}%
	\bibitem [{\citenamefont {Rao}\ and\ \citenamefont
		{M\o{}lmer}(2013)}]{Rao2013}%
	\BibitemOpen
	\bibfield  {author} {\bibinfo {author} {\bibfnamefont {D.~D.~B.}\
			\bibnamefont {Rao}}\ and\ \bibinfo {author} {\bibfnamefont {K.}~\bibnamefont
			{M\o{}lmer}},\ }\href {\doibase 10.1103/PhysRevLett.111.033606} {\bibfield
		{journal} {\bibinfo  {journal} {Phys. Rev. Lett.}\ }\textbf {\bibinfo
			{volume} {111}},\ \bibinfo {pages} {033606} (\bibinfo {year}
		{2013})}\BibitemShut {NoStop}%
	\bibitem [{\citenamefont {Rao}\ and\ \citenamefont
		{M\o{}lmer}(2014)}]{Rao2014}%
	\BibitemOpen
	\bibfield  {author} {\bibinfo {author} {\bibfnamefont {D.~D.~B.}\
			\bibnamefont {Rao}}\ and\ \bibinfo {author} {\bibfnamefont {K.}~\bibnamefont
			{M\o{}lmer}},\ }\href {\doibase 10.1103/PhysRevA.90.062319} {\bibfield
		{journal} {\bibinfo  {journal} {Phys. Rev. A}\ }\textbf {\bibinfo {volume}
			{90}},\ \bibinfo {pages} {062319} (\bibinfo {year} {2014})}\BibitemShut
	{NoStop}%
	\bibitem [{\citenamefont {Albert}\ \emph {et~al.}(2016)\citenamefont {Albert},
		\citenamefont {Bradlyn}, \citenamefont {Fraas},\ and\ \citenamefont
		{Jiang}}]{albert_adiabatic_2016}%
	\BibitemOpen
	\bibfield  {author} {\bibinfo {author} {\bibfnamefont {V.~V.}\ \bibnamefont
			{Albert}}, \bibinfo {author} {\bibfnamefont {B.}~\bibnamefont {Bradlyn}},
		\bibinfo {author} {\bibfnamefont {M.}~\bibnamefont {Fraas}}, \ and\ \bibinfo
		{author} {\bibfnamefont {L.}~\bibnamefont {Jiang}},\ }\href {\doibase
		10.1103/PhysRevX.6.041031} {\bibfield  {journal} {\bibinfo  {journal} {Phys.
				Rev. X}\ }\textbf {\bibinfo {volume} {6}},\ \bibinfo {pages} {041031}
		(\bibinfo {year} {2016})}\BibitemShut {NoStop}%
	\bibitem [{\citenamefont {Brion}\ \emph {et~al.}(2007)\citenamefont {Brion},
		\citenamefont {M\o{}lmer},\ and\ \citenamefont
		{Saffman}}]{collective_qubit_07}%
	\BibitemOpen
	\bibfield  {author} {\bibinfo {author} {\bibfnamefont {E.}~\bibnamefont
			{Brion}}, \bibinfo {author} {\bibfnamefont {K.}~\bibnamefont {M\o{}lmer}}, \
		and\ \bibinfo {author} {\bibfnamefont {M.}~\bibnamefont {Saffman}},\ }\href
	{\doibase 10.1103/PhysRevLett.99.260501} {\bibfield  {journal} {\bibinfo
			{journal} {Phys. Rev. Lett.}\ }\textbf {\bibinfo {volume} {99}},\ \bibinfo
		{pages} {260501} (\bibinfo {year} {2007})}\BibitemShut {NoStop}%
	\bibitem [{\citenamefont {Brion}\ \emph {et~al.}(2008)\citenamefont {Brion},
		\citenamefont {Pedersen}, \citenamefont {Saffman},\ and\ \citenamefont
		{M\o{}lmer}}]{collective_qubit_08}%
	\BibitemOpen
	\bibfield  {author} {\bibinfo {author} {\bibfnamefont {E.}~\bibnamefont
			{Brion}}, \bibinfo {author} {\bibfnamefont {L.~H.}\ \bibnamefont {Pedersen}},
		\bibinfo {author} {\bibfnamefont {M.}~\bibnamefont {Saffman}}, \ and\
		\bibinfo {author} {\bibfnamefont {K.}~\bibnamefont {M\o{}lmer}},\ }\href
	{\doibase 10.1103/PhysRevLett.100.110506} {\bibfield  {journal} {\bibinfo
			{journal} {Phys. Rev. Lett.}\ }\textbf {\bibinfo {volume} {100}},\ \bibinfo
		{pages} {110506} (\bibinfo {year} {2008})}\BibitemShut {NoStop}%
	\bibitem [{\citenamefont {Olmos}\ \emph {et~al.}(2011)\citenamefont {Olmos},
		\citenamefont {Li}, \citenamefont {Hofferberth},\ and\ \citenamefont
		{Lesanovsky}}]{Olmos2011}%
	\BibitemOpen
	\bibfield  {author} {\bibinfo {author} {\bibfnamefont {B.}~\bibnamefont
			{Olmos}}, \bibinfo {author} {\bibfnamefont {W.}~\bibnamefont {Li}}, \bibinfo
		{author} {\bibfnamefont {S.}~\bibnamefont {Hofferberth}}, \ and\ \bibinfo
		{author} {\bibfnamefont {I.}~\bibnamefont {Lesanovsky}},\ }\href {\doibase
		10.1103/PhysRevA.84.041607} {\bibfield  {journal} {\bibinfo  {journal} {Phys.
				Rev. A}\ }\textbf {\bibinfo {volume} {84}},\ \bibinfo {pages} {041607}
		(\bibinfo {year} {2011})}\BibitemShut {NoStop}%
	\bibitem [{\citenamefont {Kato}(1995)}]{Kato1995}%
	\BibitemOpen
	\bibfield  {author} {\bibinfo {author} {\bibfnamefont {T.}~\bibnamefont
			{Kato}},\ }\href@noop {} {\emph {\bibinfo {title} {{Perturbation Theory for
					Linear Operators}}}}\ (\bibinfo  {publisher} {Springer},\ \bibinfo {year}
	{1995})\BibitemShut {NoStop}%
\end{thebibliography}
\end{document}